\documentclass[mathleft]{an}
\pdfoutput=1
\usepackage{graphicx}
\usepackage{times}
\usepackage{natbib}
\bibpunct{(}{)}{;}{a}{}{,}

\everymath{\displaystyle}

\begin{document}

\Pagespan{1}{}
\Yearpublication{2014}%
\Yearsubmission{2014}%
\Month{1}%
\Volume1%
\Issue{1}%

\title{Dynamical evolution of two associated galactic bars}
\author{F. Garz\'on\inst{1,2}\fnmsep\thanks{Corresponding author: \email{fgl@iac.es}\newline} 
 \and M. L\'opez-Corredoira\inst{1,2}}
\titlerunning{Dynamics of two bars}
\authorrunning{F. Garz\'on \& M. L\'opez-Corredoira}
\institute{Instituto de Astrof\'{\i}sica de Canarias, E--38200, La Laguna (S.C. Tenerife), Spain
\and
Departamento de Astrof\'{\i}sica, Universidad de La Laguna, E--38206, La Laguna (S.C. Tenerife), Spain}

\received{XX mmm YYYY}
\accepted{XX mmm YYYY}
\publonline{later}


\abstract{
We study the dynamical interactions of mass systems in equilibrium under their own gravity 
that mutually exert and experience gravitational forces. The method we employ is to model 
the dynamical evolution of two isolated bars, hosted within the same galactic system, under
their mutual gravitational interaction. In this study we present an analytical treatment of
 the secular evolution of two bars that oscillate with respect one another. Two cases of 
interaction, with and without geometrical deformation, are discussed. In the latter case, 
the bars are described as modified Jacobi ellipsoids. These triaxial systems are formed by 
a rotating fluid mass in gravitational equilibrium with its own rotational velocity and the 
gravitational field of the other bar. The governing equation for the variation of their 
relative angular separation is then numerically integrated, which also provides the time
evolution of the geometrical parameters of the bodies. The case of rigid, non-deformable, 
bars produces in some cases an oscillatory motion in the bodies similar to that of a harmonic 
oscillator. For the other case, a deformable rotating body that can be represented by a
modified Jacobi ellipsoid under the influence of an exterior massive body will change its 
rotational velocity to escape from the attracting body, just as if the gravitational torque 
exerted by the exterior body were of opposite sign. Instead, the exchange of angular momentum
 will cause the Jacobian body to modify its geometry by enlarging its long axis, located in
 the plane of rotation, thus decreasing its axial ratios.}

\maketitle

\section{Introduction}
\label{.intro}

The dynamical interactions between different galactic components is of growing interest
 with regard to modelling the self-consistent evolution of a stellar system, since it may provide 
important clues on the formation and environmental conditions of such a system. Within this 
general subject, many attempts can be found in the literature to investigate the dynamics of a 
barred galactic system from several different standpoints. \citet{Abramyan1986} and \citet{Louis1988} 
made the first trials to provide a theoretical framework within which the mutual evolution of 
two triaxial systems, namely the bar and bulge components of barred galaxies \citep{Louis1988} 
combined with other components \citep{Abramyan1986}, can be followed. The work by \citet{Louis1988},
based on Jeans's theorem in extended phase space, provided some insights into oscillating 
systems governed by a periodically time-dependent gravitational potential, while the paper of \citet{Abramyan1986} 
considers the equilibrium conditions for both collisionless and fluid-dynamical models of a 
self-gravitating bulge within a complex system of bar, disc and halo. \citet{MartinezValpuesta2006} 
have analysed the evolution of a galactic bar, focusing on the vertical buckling instability that 
would lead to pronounced changes in the shape of the bulge, which would then appear as 
boxy/peanut-shaped giving rise to a pseudo double bar system; \citet{Weinberg2007,Weinberg2007a} 
have investigated the dynamics of bar--halo interactions using perturbation theory and numerical 
experiments; \citet{Maciejewski2007, Maciejewski2008} have analysed the orbits and trajectories of 
particles which form the backbone of nested bar systems; \citet{Debattista2007} (hereafter DS07) 
and \citet{Shen2009} consider the case of a barred galaxy hosting a secondary bar. These authors, 
and others, have made substantial contributions to the setting of the problems, using N-body 
simulations in most cases, although the case of mutual interaction of two bars has not yet been 
a subject of study.

In this contribution we summarize the first results of an ongoing investigation into the matter 
with the aim at providing a new theoretical approach to the problem of dynamical interaction and
 evolution in stellar systems. We have purposely selected the case of double-bar systems given
 the increasing amount of observational evidence of their high relative frequency in the Universe
 \citep[see][and references therein, although this review focuses on galaxies where a smaller 
secondary bar is nested inside a larger primary bar]{Erwin2008}, the different models available 
for their treatment and the evidence for a double-bar structure in the central region of the Milky
 Way. To this end, we have followed a purely analytical approach using the basic equations of  
classical mechanics and the treatment of ellipsoidal bodies in equilibrium with their own gravity 
\citep{Lyttleton1953, Chandrasekhar1969}.

It should be noted that our own galaxy, the Milky Way, most probably hosts a double-bar type 
structure at its centre---the long thin bar and the boxy bulge, whose detailed nature and very 
existence are the subject of some controversy. The rationale for the Milky Way being a barred 
galaxy has steadily gained acceptance following many discoveries concerning the morphology of the
 central regions of the Galaxy, mostly from detailed NIR star counts. The first evidence for a 
bar-like distribution of the stars was derived from the asymmetries in the infrared surface brightness
 maps \citep[e.g.][]{Matsumoto1982,Blitz1991,Dwek1995} and in source counts \citep{Weinberg1992,
 Hammersley1994, Stanek1994}, which all
 show systematically more stars at positive galactic 
longitudes within $\ell < 30^\circ$ and close to the Galactic plane compared to negative longitudes. 
The exact morphology of the inner Galaxy, however, is still controversial. While some authors 
refer to a fat structure, around 2.5 kpc in length with a position angle (PA) of 15\degr --\,30\degr 
with respect to the Sun--Galactic Centre direction \citep{Dwek1995, Nikolaev1997, Stanek1997, 
Binney1997, Freudenreich1998, Bissantz2002, Babusiaux2005}, other researchers suggest that there 
is also a long thin bar, the in-plane bar, with a half-length of 4 kpc and a position angle of 
around $45^\circ$ \citep{Peters1975, Weinberg1992, Stanek1994, Benjamin2005}. It is noteworthy 
that those authors supporting the $23^\circ$ PA bar all examine the region at $|\ell|< 12^\circ$, 
whereas those supporting the long bar with the larger PA are trying to explain counts for 
$10^\circ < |\ell| < 30^\circ$. It is also useful to mention that several authors in the
 first group assume that it is the boxy bulge of the central Galaxy which produces the bar-like 
structure seen in the observational data. This variety of standpoints is also present for the nature of the double bar structure in external galaxies possessing this feature. In \citet{Compere2014}, we present some evidences for two different bar-like morphologies within the same galaxy.

The detailed geometrical morphology and luminosity 
function of this boxy bulge was delineated in \citet{Lopez-Corredoira2005} by Bayesian inversion 
of the stellar statistics equation using 2MASS star counts. Analysing the TMGS $K$                                                                                                                                                                                                                                                                                                                                                                                                                                                                                                                                                                                                                                                                                                                                                                                                                                                                                                                                                                                         -band star counts, 
\citet{Hammersley1994} and \citet{Calbet1995} initially posited the existence of a long bar of 
radius 4 kpc, with its closest tip at $\ell=27^\circ$ and the farthest at $\ell=-22^\circ$, thus
 indicating a PA of of $75^\circ$. \citet{Calbet1996} claimed the existence of a dust lane preceding 
the bar at negative longitudes as expected for a rotating bar, which would explain the higher 
extinction observed in this region. The analysis of the TMGS database resulted in spectroscopic 
follow-up of selected samples. Thus \citet{Garzon1997} and \citet{Lopez-Corredoira1999} reported
 a large excess of supergiants in the $\ell=27^\circ$ area, covered by the TMGS. Such regions form
 because of the concentrations of shocked gas where a galactic bar meets a spiral arm, as observed
 at the ends of the bars of face-on external galaxies. Thus, the presence of a massive star formation
 region is very strong supporting evidence for the presence of a bar in our Galaxy. 
\citet{Hammersley2000} made use of a preliminary data set of the TCS--CAIN survey in which the red-clump
 stellar population could be identified and then used as a standard candle for distance calculation. 
In so doing, it became clear to them that the geometry of the bar needed to be re-estimated to yield 
a revised PA of $43^\circ$, after correcting the position of the farthest end of the bar, 
now at $\ell=-12^\circ$. New data analyses are also shown in \citet{Lopez-Corredoira2007a} and 
\citet{Cabrera-Lavers2007}.


\begin{figure}[h]
\resizebox{\hsize}{!}{\includegraphics[angle=0,width=0.4\textwidth]{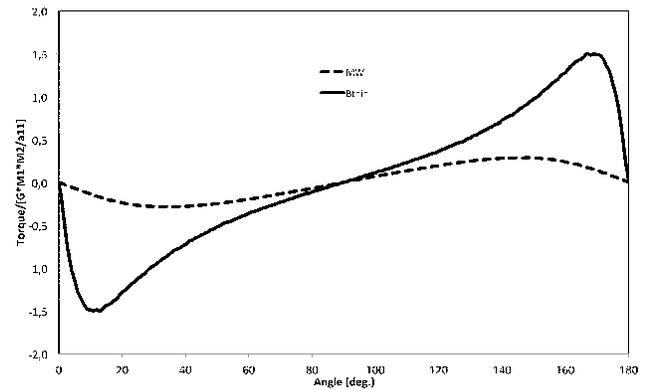}}
\caption{Variation of the torque with the angle between bars. Two cases are shown: solid line, 
labelled Bthin, is for two identical bars with $q_2=q_3=0.1$; and dashed line, labelled MW, 
represents a configuration similar to that of the Milky Way (see text), with $q_2=0.22$ and $q_3=0.04$, 
for the long bar, and $q_2=0.5$ and $q_3=0.4$, for the boxy bulge with its long axis half the length of 
that of the long bar. The torque is given in units of $G\times M_1\times M_2/ a_{11}$, $a_{11}$ being 
the semi-axis of the first bar. The ripples in the graphs are due inaccuracies of the numerical 
integration method (see text).} 
\label{fig_torq}
\end{figure}

It seems tf1.imely to investigate from a theoretical point of view the nature of the interactions 
between two bar-like bodies that coexist within the same area of a galaxy like ours, and this is 
the main objective of this paper. We start, as is customary, by analysing the problem focusing on 
the mutual interaction of the two bars isolated from the rest of the system. It is our purpose to 
extend the reach of this research by including additional material structures from the host galaxy 
(the stellar and interstellar disc, halo, etc.) in subsequent phases, which will be reported in due course.

\section{The study case}
The general gravitational interaction between two rotating bodies can be described via the interchange 
of angular momentum, governed by the run of the mechanical mutual torque exerted on both bodies.
In this analysis, we follow the secular evolution of two isolated triaxial bars under their mutual 
influence. We assume, for the sake of simplicity, that both bars are of uniform density, not necessarily 
equal, rotate about the same single axis and through the same pivotal point and can be represented by 
triaxial ellipsoids, each with its own geometric parameters. Each bar can then be regarded as a solid 
body comprising a stellar population orbiting in trapped trajectories within each bar potential, 
which will be scaled to the geometry of the bar, always keeping the same figure. The only 
external interaction considered  on each bar will be the gravitational torque created by the other bar.

In this way, we can write the equations for the run of both the torque, $\vec{\tau}$, and angular
momentum, $\mathbf{L}$, as

\begin{equation}
{\setlength\arraycolsep{0.5em}
\begin{array} {ccccccc}
\vec{\tau}_\mathrm{T} = \vec{\tau}_1 + \vec{\tau}_2 = 0 &  &
\vec{\tau}_\mathrm{T} = {d\mathbf{L}_\mathrm{T}}/{dt} \\
\vec{\tau}_i = {d\mathbf{L}_i}/{dt} &  &
\mathbf{L}_i= \bar{\bar{I_i}} \mathbf{\Omega}_i
\end{array} 
}
\label{eq_ini}
\end{equation}
the magnitudes with the subscript `T' pertaining to the system as a whole, and those with {\it i} 
to each of the bars. $\bar{\bar{I_i}}$ is the tensor of inertia of each bar and $\mathbf{\Omega _i}$, 
the angular velocity. Let us set the $Z$-axis as the axis of rotation so that the only component of 
interest for the intervening magnitudes is the third one, along the $Z$-axis, and the torque, angular
 momentum and angular velocities can be treated from now on as scalars, considering their components
 along the $Z$-axis only. The moment of inertia of interest is then the third component of the diagonal 
tensor of inertia written in the reference system of the ellipsoid principal axes.

\begin{figure*}
  $\begin{array}{lr}
  \includegraphics[angle=-90,width=0.49\textwidth]{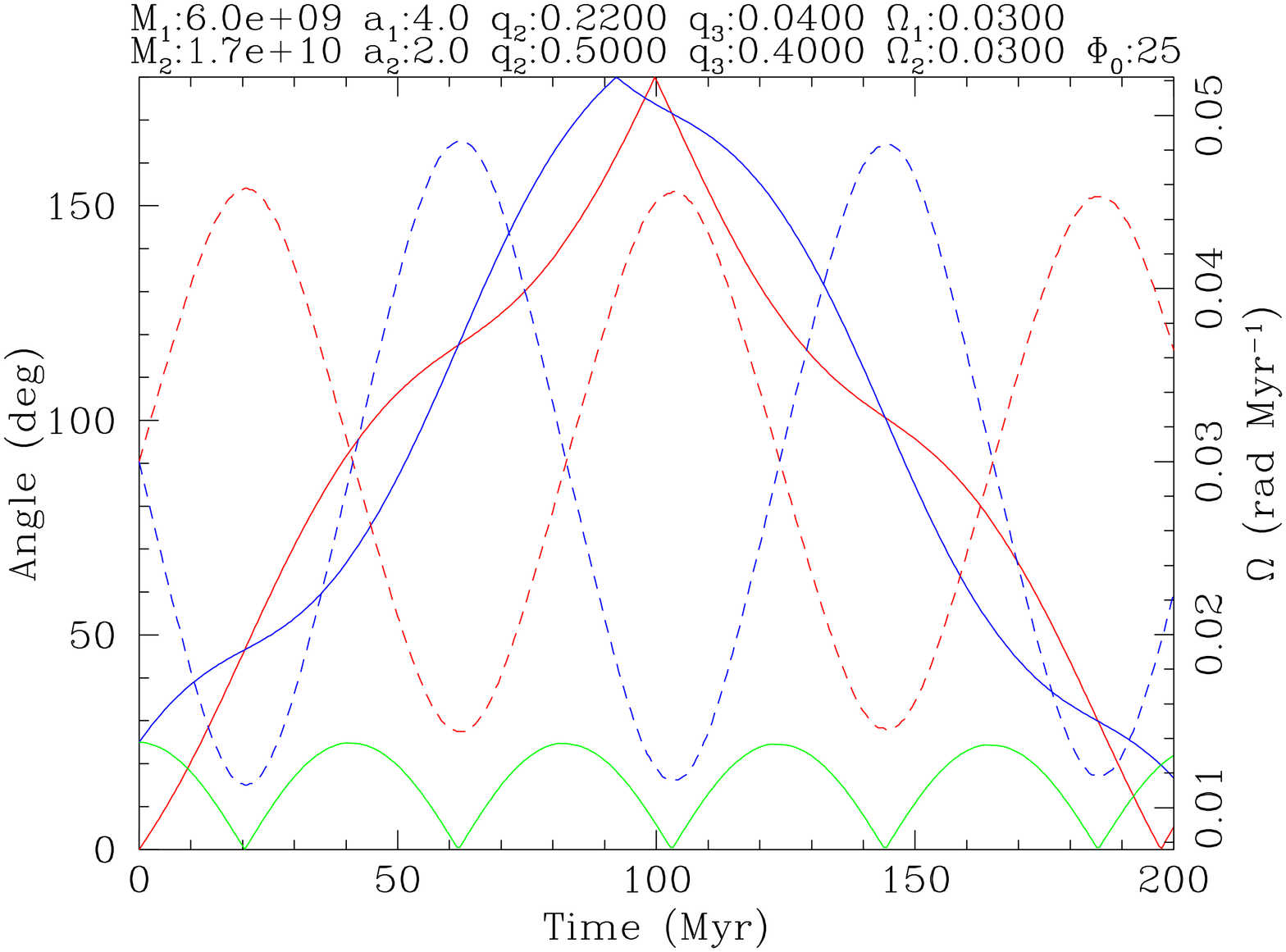} &
 \includegraphics[angle=-90,width=0.49\textwidth]{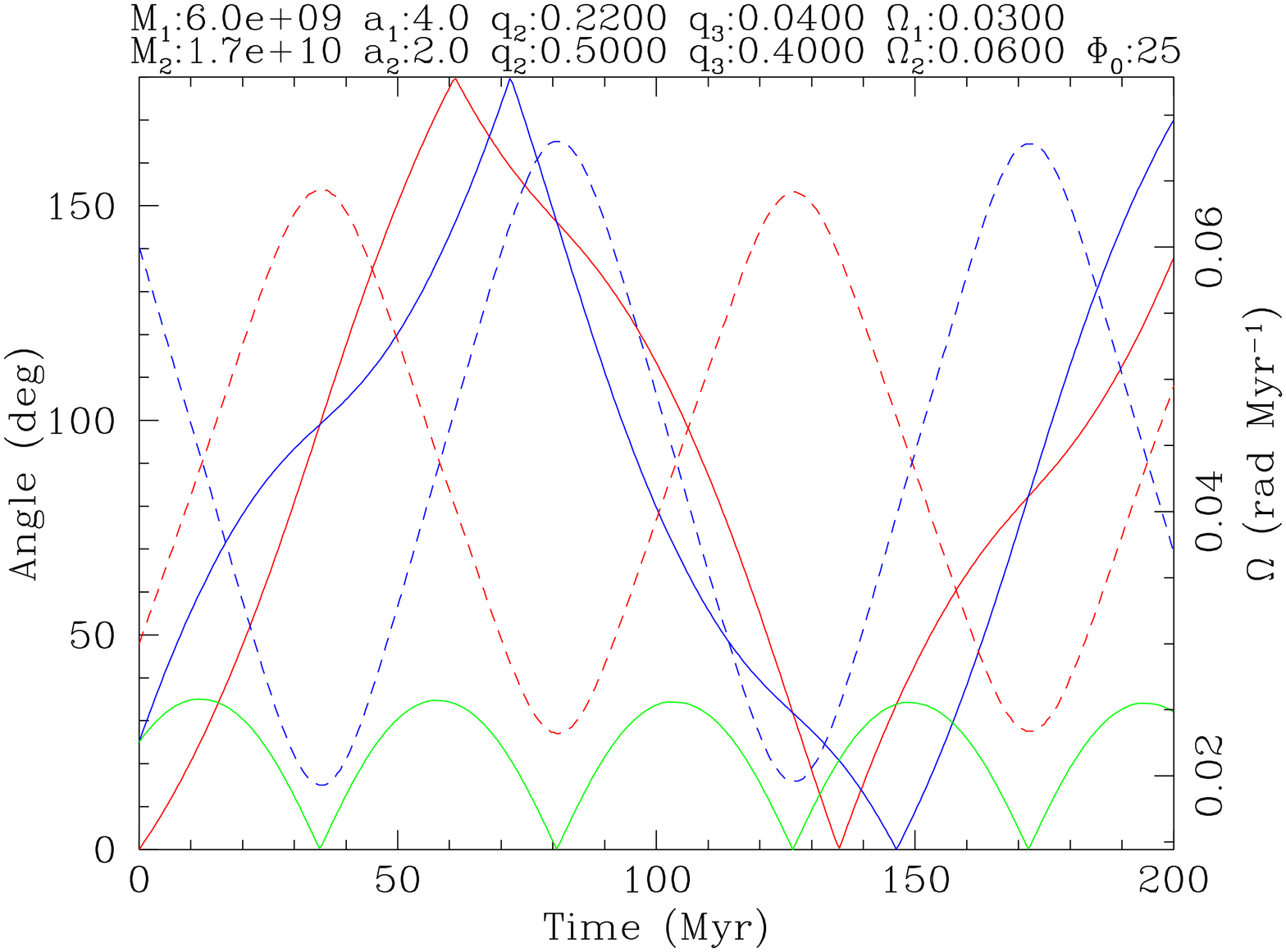} \\
  \includegraphics[angle=-90,width=0.49\textwidth]{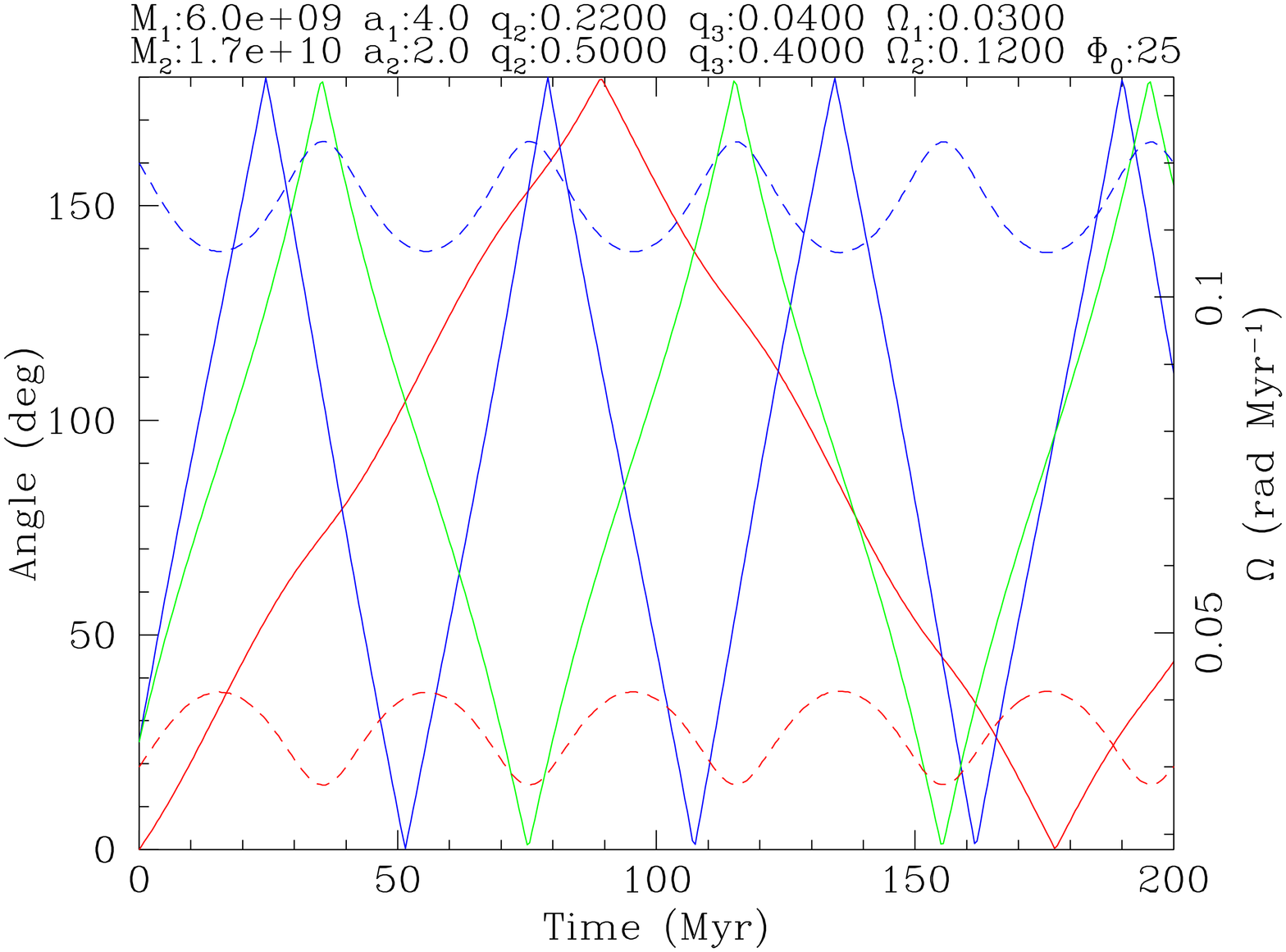} &
  \includegraphics[angle=-90,width=0.49\textwidth]{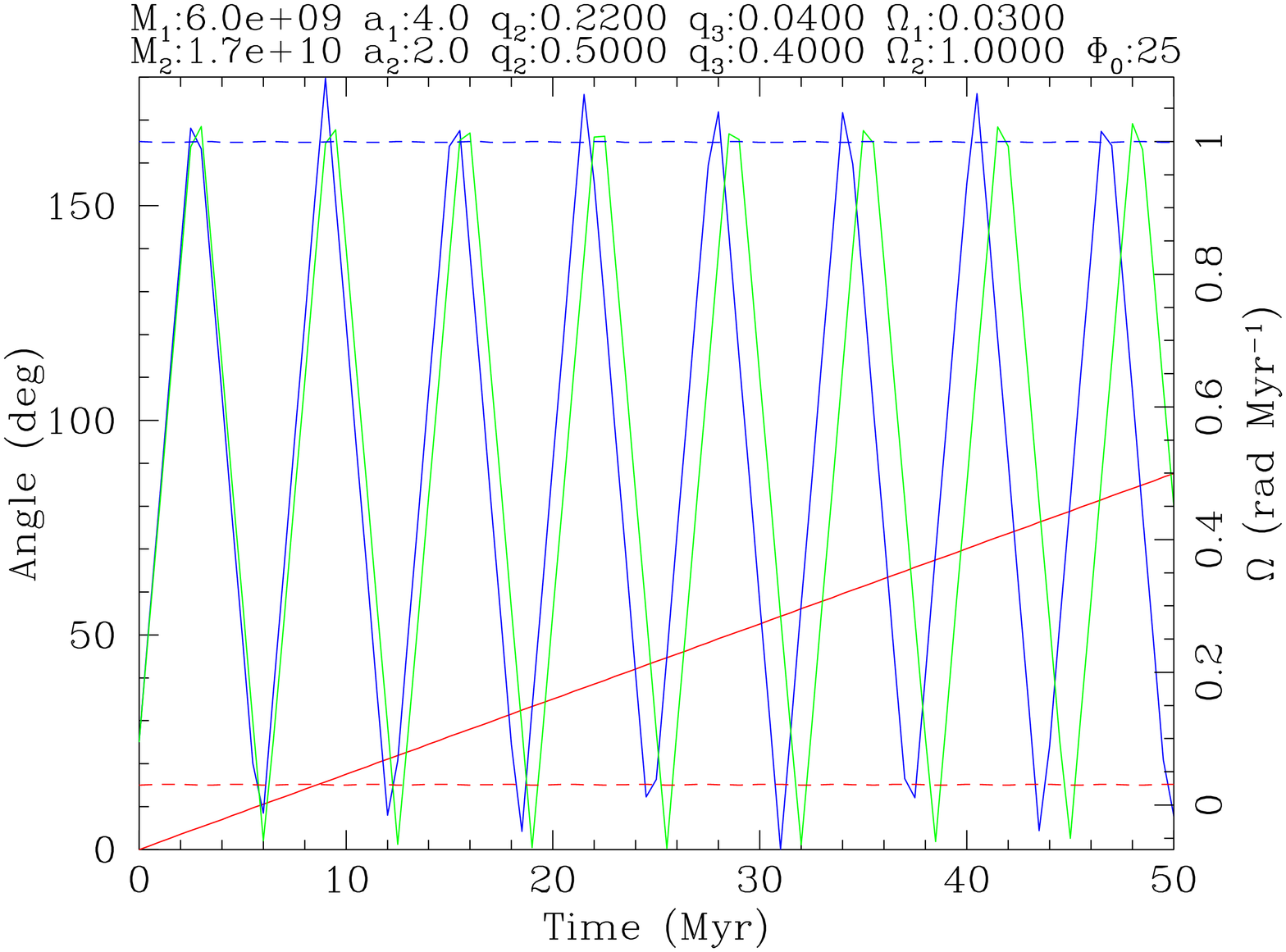} \\
  \includegraphics[angle=-90,width=0.49\textwidth]{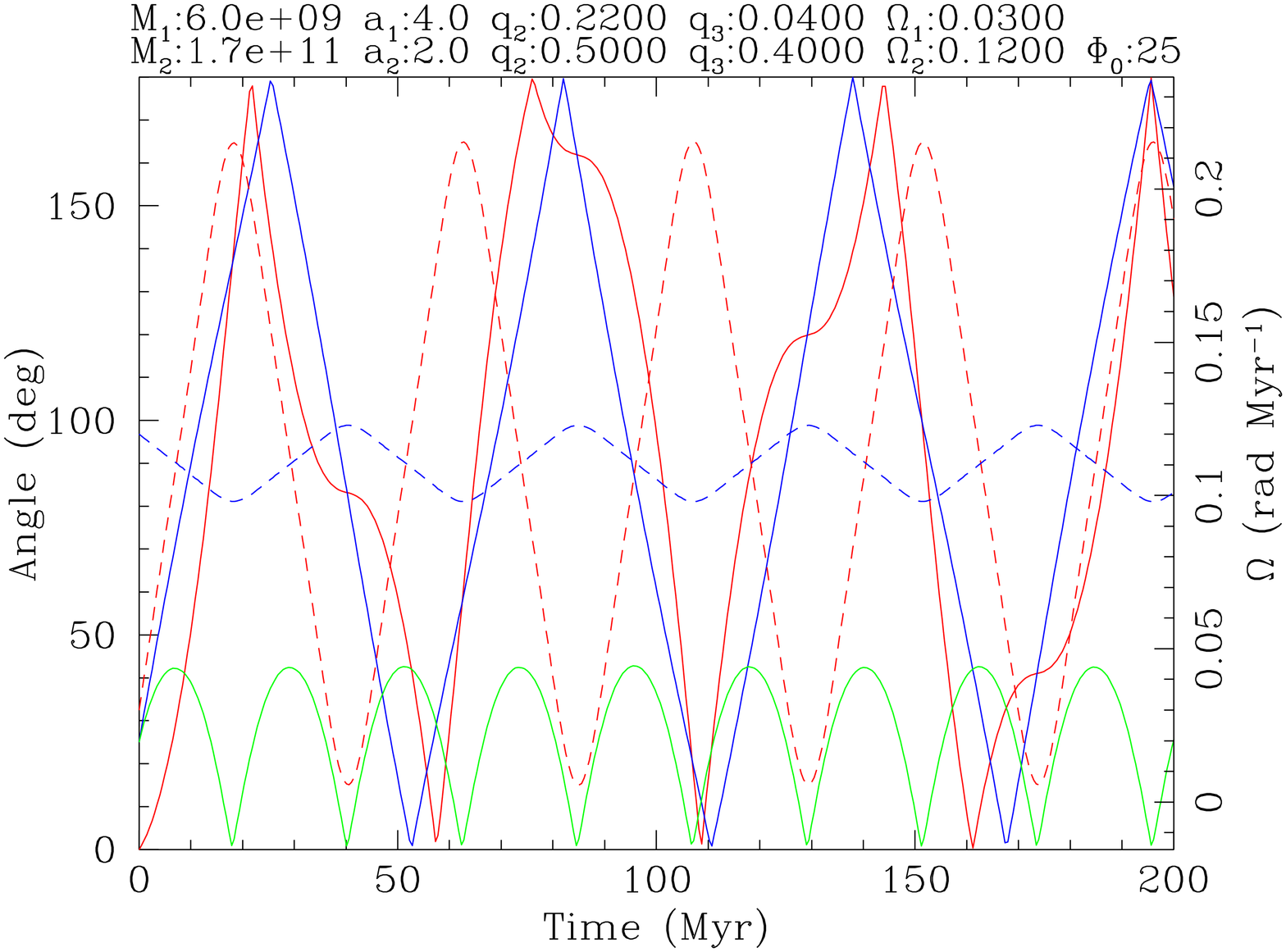} &
 \includegraphics[angle=-90,width=0.49\textwidth]{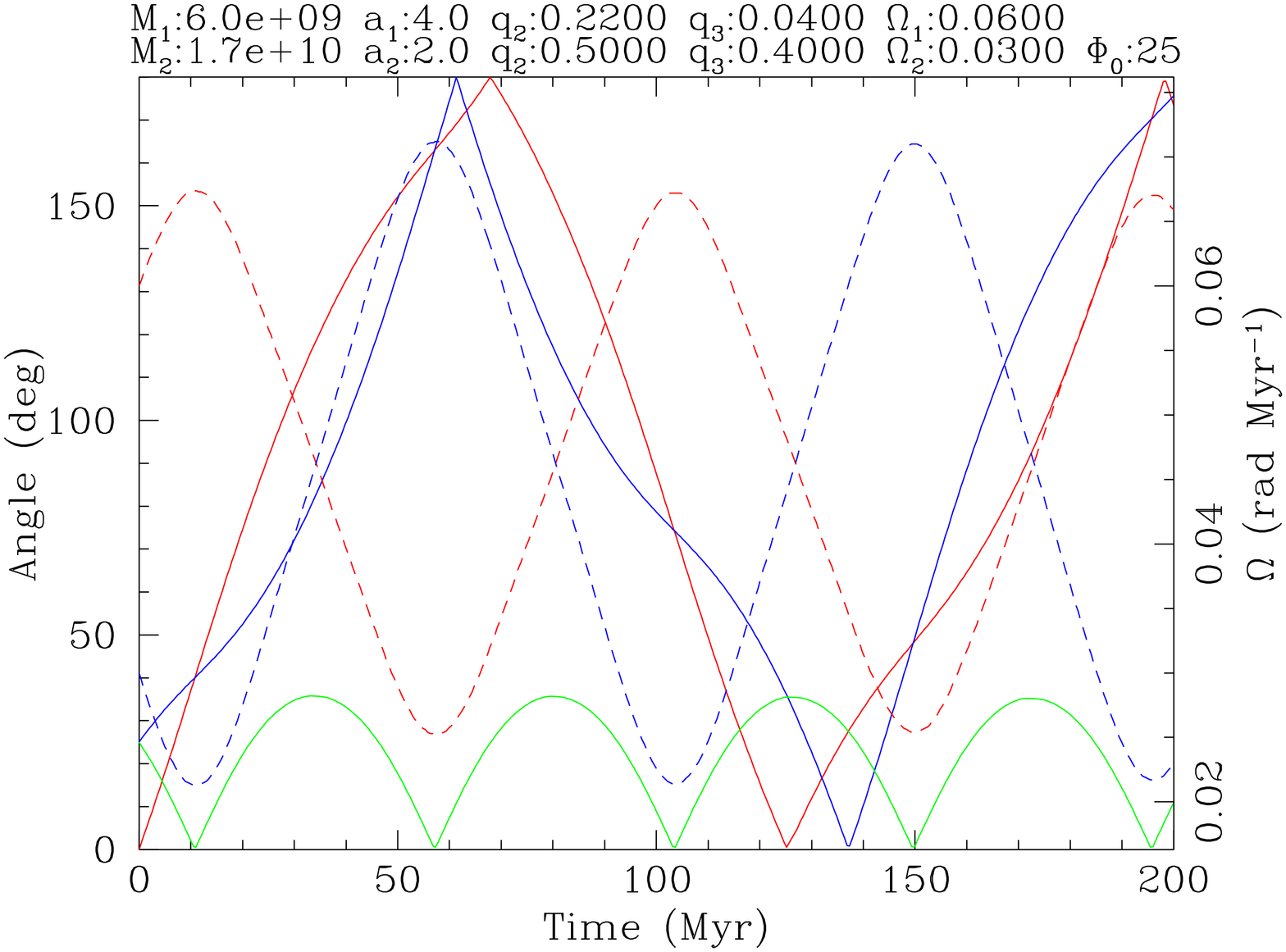} 
  \end{array}$
\caption{Evolution with time of several geometrical and dynamical parameters of two rigid bars for six different configurations. At the top of each panel the initial parameter set for each bar is given. $\Phi_0$ is the initial angle between bars, assuming always that bar 1 lies along the $X$-axis at the start. Units for the parameters are given in the text. Solid lines show the PAs, in red for bar 1 and in blue for bar 2; the green line is the relative angle between bars. Angles are plotted in the range [0\degr,180\degr] by replacing PA by 360\degr-PA for PA $>$ 180\degr. Dashed lines represent the angular velocities for each bar, with the same colour code, plotted against the secondary $Y$-axis.}
\label{fig_nj}
\end{figure*}

\begin{figure}[h]
{\par\centering \includegraphics[width=0.48\textwidth]{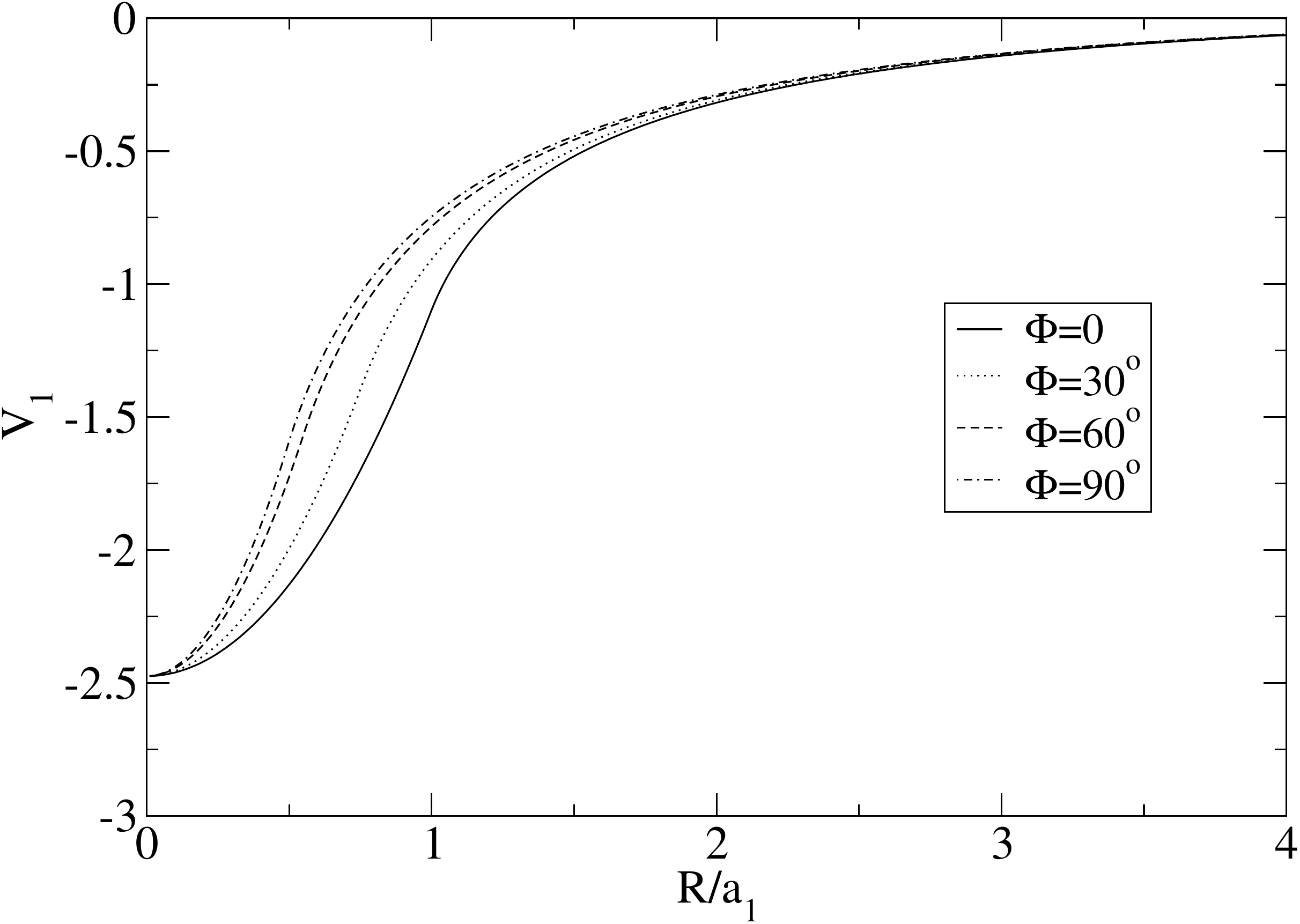}\par}
\caption{Potential created in the plane by a homogeneous ellipsoid with $q_2=0.5$, $q_3=0.25$ and 
semimajor axis $a_1$, following Eq. (\protect{\ref{potencial}}). Units are such that $GM/a_1=1$.}
\label{Fig:potbar}
\end{figure}

Hence, the governing equation for the evolution of the system is

\begin{equation}
{\setlength\arraycolsep{0.2em}
\begin{array} {lll}
 \ddot{\Delta\Phi} &=& \left(\dot\Omega_2 - \dot\Omega_1 \right) \\
 &=&\left( {\frac{1}{{I_2 }} + \frac{1}{{I_1 }}} \right)\tau _2 (\Delta \Phi ) - \frac{{\mathrm{d}ln(I_2 )}}{{\mathrm{d}t}}\Omega _2 +\\
 &&+ \frac{{\mathrm{d}ln(I_1 )}}{{\mathrm{d}t}}\Omega _1
 \end{array}}
\label{eq_ppal}
\end{equation}
which gives the variation of the relative position angle of the bars, $\Delta \Phi$ , 
in terms of the moment of inertia, $I_i$, angular velocities, $\Omega _i$, and gravitational torque, 
$\tau_i (=-\tau_{j\neq i})$. The two scalar moments of inertia, $I_1$ and $I_2$, are given by

\begin{equation}
I_i = \frac{1}{5}M_i (a_{1i}^2 + a_{2i}^2 ) = \frac{1}{5}M_i a_{1i}^2 (1 + q_{2i}^2 )
\label{eq_moi}
\end{equation}
$a_{1i}$ and $a_{2i}$ being the semi-axes of the ellipses in the equatorial plane ($XY$) and the axial
 ratios with respect to $a_{1i}, q_{2i}\equiv a_{2i}/ a_{1i}$. The parameters $q_{3i}$, which will
 appear later, are also defined in this way, as the third axial ratios with respect to $a_{1i}, a_{3i}/ a_{1i}$.

The torque, its component along the $Z$-axis, can be expressed as a function of the azimuthal component 
of the force exerted by each elemental mass volume in bar $i$ to the bar $j$ at $\mathbf{r}, 
F_{\Phi i}$, \citep[see eq. (6) in][]{Lopez-Corredoira2007} as:

\begin{equation}
\tau _j (\Delta \Phi ) = \int_{V_j} {\mathrm{d}V\rho_j } F_{\Phi,i} (\mathbf{r})R
\label{eq_torq}
\end{equation}
where $R$ is the radial component of the position vector $\mathbf{r}$ in the cylindrical coordinate 
system. Equation \ref{eq_torq} is solved numerically for each of the study cases using a composite 
Simpson rule for multiple integrals. Details of the treatment of Eq.\ (\ref{eq_torq}) are given in 
Appendix \ref{app_torq}. It needs to be stated, however, that the calculation of this multivariate 
integral using a rather simple quadrature produces small artificial fluctuations in the data of the 
type that can be seen in Figure \ref{fig_torq} . This introduces small numerical errors at every 
integration step that can accumulate throughout the entire process. Thus, it is important to make the step 
length of the quadrature small enough to increase the accuracy of the result. This is even more crucial 
in ranges where the integrand experiences steeper variations. We have taken these actions into account in 
the integration.

In what follows we present some examples of the results obtained so far by integrating Eq.\ (\ref{eq_ppal}) 
in the two cases of rigid rotation (i.e.\ no variation in the moment of inertia with time is allowed) and
 deformable rotation, in which the moment of inertia varies with time. In both cases, the bars are assumed 
to be of uniform density. The case for heterogeneous bodies, under certain prescriptions, will be developed 
in a subsequent study.

\subsection{The case for rigid rotation}
\label{.srig}

This is the simplest case and can be studied quite easily with the help of Eq.\ (\ref{eq_ppal}), in 
which the last two terms have been dropped. A
multistep method is then implemented by sequentially incrementing the time to follow the evolution of the double bar system.
As initial parameters for the integration to proceed, we have selected the following:
\begin{itemize}
\item The masses, $M_1$ and $M_2$, of the bars, which remain unaltered in the integration.
\item The lengths of the semi-axes, $(a_1,a_2,a_3)$, of each bar, which are also unaltered during
 the evolution of the system.
\item The initial angle, $\Delta \Phi$, between the major axis of bar 2 with respect to that of bar 1, 
and the initial rotational velocities of each bar.
\end{itemize}
The evolution of the system is controlled by Eq.\ (\ref{eq_ppal}), which can be rewritten in this case as

\begin{equation}
 \ddot{\Delta\Phi} = \left(\dot\Omega_2 - \dot\Omega_1 \right) = \left( {\frac{1}{{I_2 }} + \frac{1}{{I_1 }}} \right)\tau _2 (\Delta \Phi )
\label{eq_ppal_rig}
\end{equation}
and $\tau_2$ is computed for each relative angle following the prescriptions in Appendix \ref{app_torq}.
 
Given the initial parameters already detailed, the integration can now proceed. We show results for a 
parameter configuration representing the likely configuration of the central zone of the Milky Way, 
which comprises a long thin bar and a thick boxy bulge, whose geometries have been selected according 
to the structural parameters derived for the Galactic long bar and bulge \citep{Lopez-Corredoira2005, 
Lopez-Corredoira2007a, Sevenster1999}: $M_1=6\times10^9, M_2=1.7\times10^{10}; a_{11}=4, q_{21}=0.22, 
q_{31}=0.04; a_{12}=2, q_{22}=0.5, q_{32}=0.4$; and with angular frequencies in the range $\Omega_1,\Omega_2\in[0.03,1]$. 
Units are solar masses, kpc, rad Myr$^{-1}$, and sexagesimal degrees. We have also explored departures 
from this parameter space, part of which will be also shown here.

The evolution of the system is solely controlled by the variation of the torque with the angular 
separation of the bars. See Fig.\ \ref{fig_torq} for the run of the torque for the case of two bars
 with a parameter set given in the caption. It has to be noted that the torque between the two bars 
is independent of these mass distributions being rigid or deformable and reflects only the action 
between the two distributions with a given configuration. The overall run of the torque is maintained 
in all the configurations, which does not depend on the angular speed of the bars. The angle at which 
the torque peaks is a function of the relative thickness of the two bars. It can be seen that the peak
 angle, with different sign at each side, roughly equals  the angle at which the tip of the shortest bar, 
sketched for this purpose solely as a rectangle of half sides $a_{1i}$ and $q_{2i}\times a_{1i}$, abandons
 the location of the longest one, also represented in the same way for the calculation.

Results for this case of non-deformable bars are shown in Fig.\ \ref{fig_nj} for six cases with mass ratios
 $M_1/M_2\approx[1/3,1/30]$, and $\Omega_1,\Omega_2=[0.03,0.06,0.12,1.0]$. For the cases where the angular 
speed is similar, panels 1, 2 and 6 numbered from top to bottom and left to right, the lightest bar (bar 1 
in the plots) remains trapped and oscillating around the more massive bar (bar 2), as also noted in 
\cite{Heller2001a} but for the secondary nuclear gaseous bar. In these trapped cases the oscillation 
pattern remains the same. Bar 1 experiences the largest amplitude in variation of angular velocity, 
which increases with time. Bar 2 follows a similar pattern but at reduced amplitude. The amplitude in the 
oscillation of the relative angle also increases with time until the bars become coupled on the opposite side,
 when the overall motion pattern is repeated. Without these similar angular velocities, it is interesting
 to see how case 5 becomes a bounded system, in terms of developing oscillations, in contrast to case 3, 
sharing both the same frequencies of 0.03 and 0.12 rad Myr$^{-1}$ for bars 1 and 2. In the former, the 
mass of bar 2 is 10 times higher than in the latter case, which is sufficient to trap bar 1.

For the computation to produce trustworthy results, it is important to keep the time step within a 
tolerable range over which the intervening magnitudes in Eq.\ (\ref{eq_ppal_rig}) can be considered as 
constant. This requires the number of integration steps to be very large to follow the evolution of
 the system over a sufficient number of oscillation periods. Since, as commented in Appendix \ref{app_torq}, 
we have selected a simplified method for solving the integral equation that gives the torque, we can 
decrease the size of the time step while keeping the computational time within reasonable limits. Throughout this paper, the time 
step has been selected in the range [0.1,0.5] Myr, which we have tested produces reliable results, 
without the need of implementing further corrective actions.

This rigid bar case hence produces a fixed oscillation pattern that lasts forever, as some sort of periodic 
oscillator, in the absence of external action on the system. This is a direct consequence of the 
conservation of both total angular momentum and energy that governs the evolution of the system as 
is developed in full in Appendix \ref{app_per}.

The amplitude of the oscillations in angular velocity, and consequently in relative angle, depends on the
 mass ratios, being bigger for each bar the smaller this ratio is, and not on which bar rotates more rapidly,
 as can be seen by comparing the top right and bottom right panels: bar 1 is faster than bar 2, 
and the motion pattern is very much the same. The time period for cases 1, 2, 5 and 6 represented in
Fig.\ \ref{fig_nj} are 82.4, 91.1, 44.5 and 91.9 Myr, this period being the time length for the relative
 angle to perform a full oscillation. Examples 3 and 4 do not develop a periodic motion since the bars
 move at very different speeds and the mutual action does not significantly change the initial motion patterns.

\subsection{The case for deformable rotation}
\label{.sdef}

To account for the variation of the moment of inertia, which follows the geometrical deformation 
of the bar with the interaction and time, the shape of the two bars has to be assigned to a
given equilibrium figure. Triaxial ellipsoids seem to be the natural choice given the observed 
morphology of bars in external galaxies of a variety of orientations. Should the bars be isolated 
they would have been modelled as Jacobi ellipsoids, rotating fluid bodies in gravitational equilibrium
 with their own rotational velocity \citep{Lyttleton1953, Chandrasekhar1969}. Jacobi ellipsoids are
 homogeneous bodies and have a unique relation between the ratios of the axes,
 $q_2=a_2/a_1$ and $q_3=a_3/a_1$, being $(a_1\ a_2\ a_3)$ the semi-axes of the bounding ellipsoids 
describing the bars, for the equilibrium to be at all possible, with which a unique angular velocity 
is associated for each figure. But in our case the bars form a coupled system in which each of the 
two must be in equilibrium not only with its own gravity but with the combined action of its own and the 
the other bar's gravitational action. In this section we will develop an analytical approach, 
following closely that of the Jacobi ellipsoids, to derive the total gravitational potential acting 
on each bar and the resulting equations for the axes and rotational velocities of each bar. At this 
stage the bars will be considered as of homogeneous densities, but not as necessarily equal to each
other.

\begin{figure}
  $\begin{array}{c}
  \includegraphics[angle=0,width=0.47\textwidth]{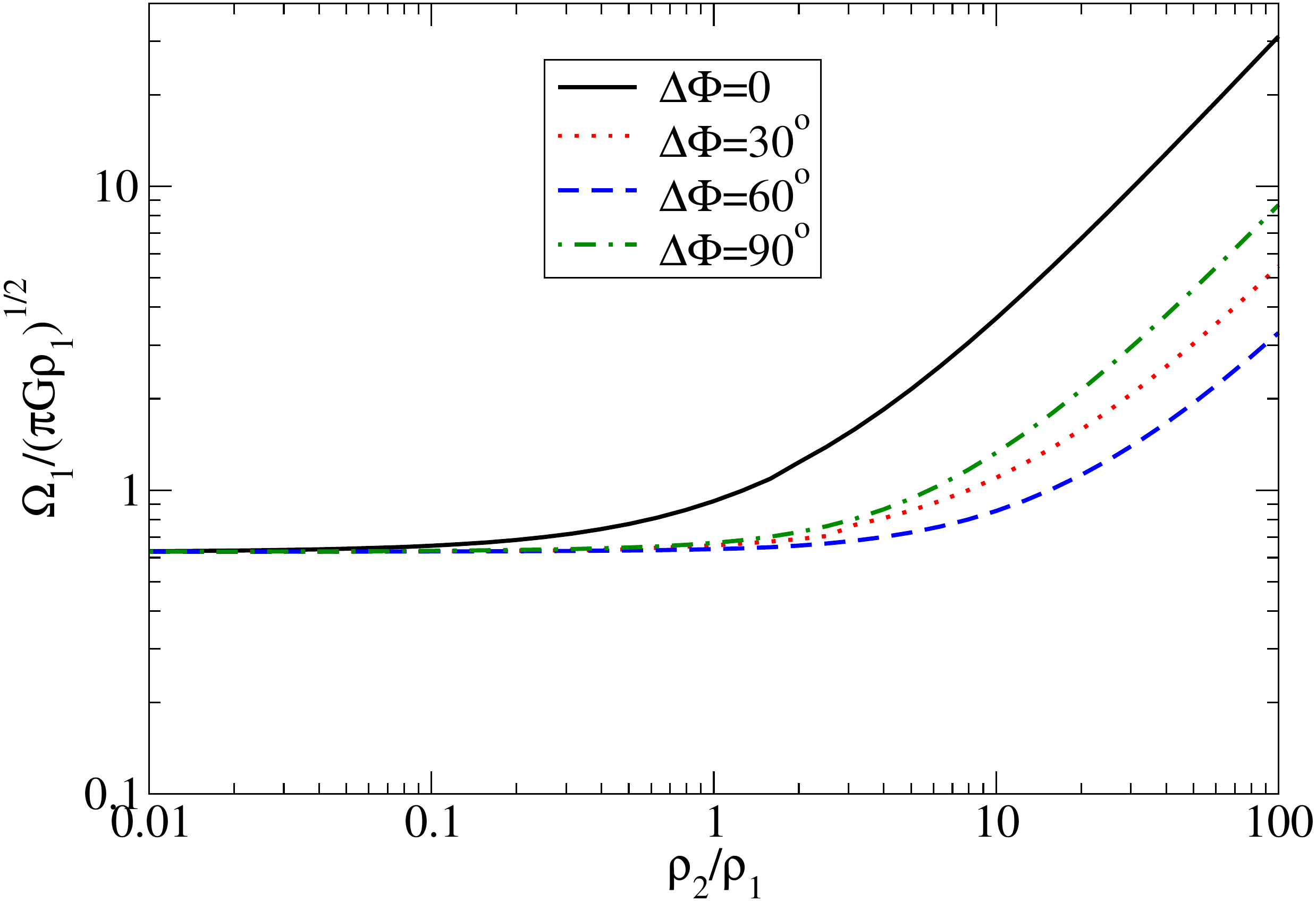} \\
  \includegraphics[angle=0,width=0.47\textwidth]{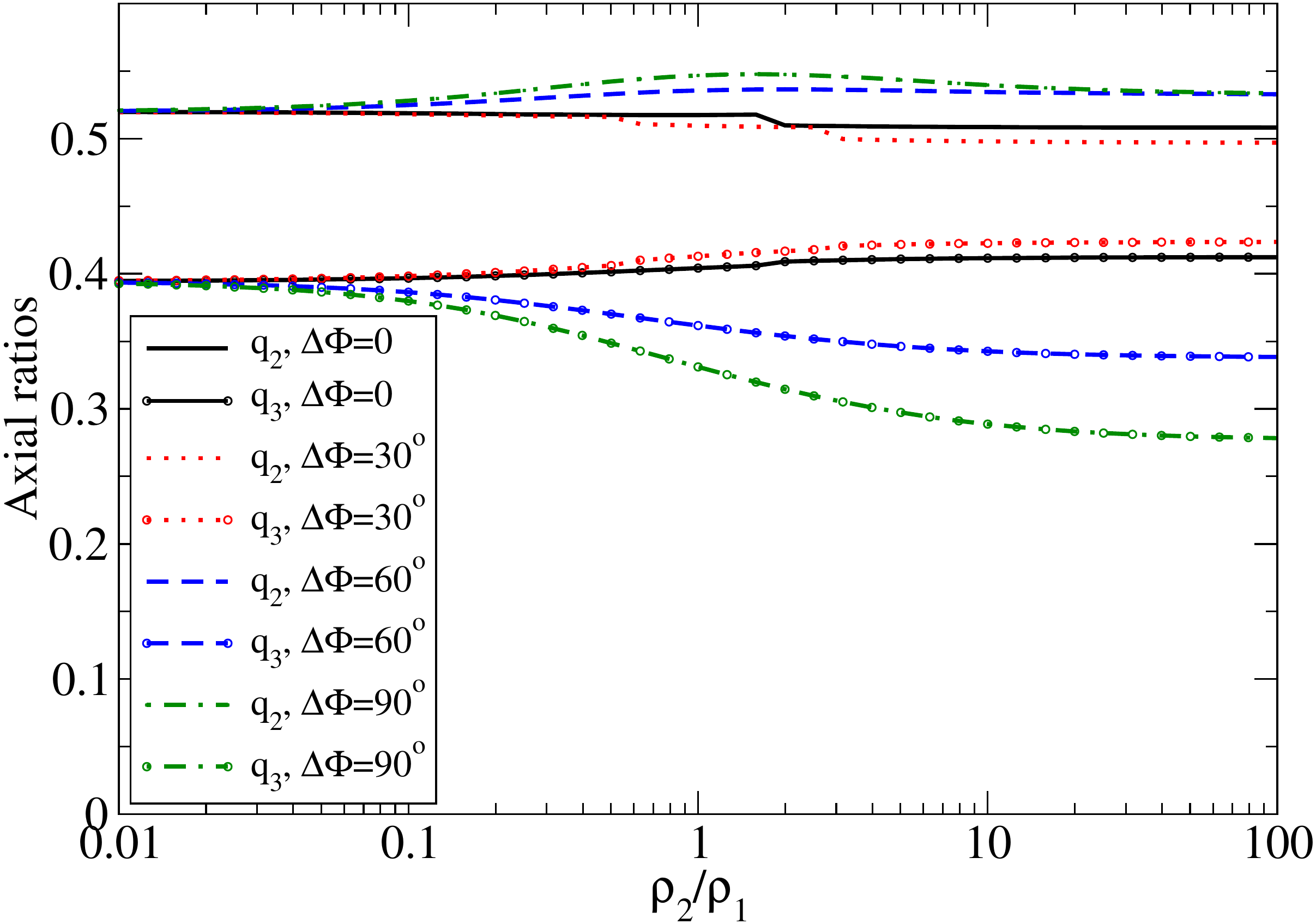} \\
  \includegraphics[angle=0,width=0.47\textwidth]{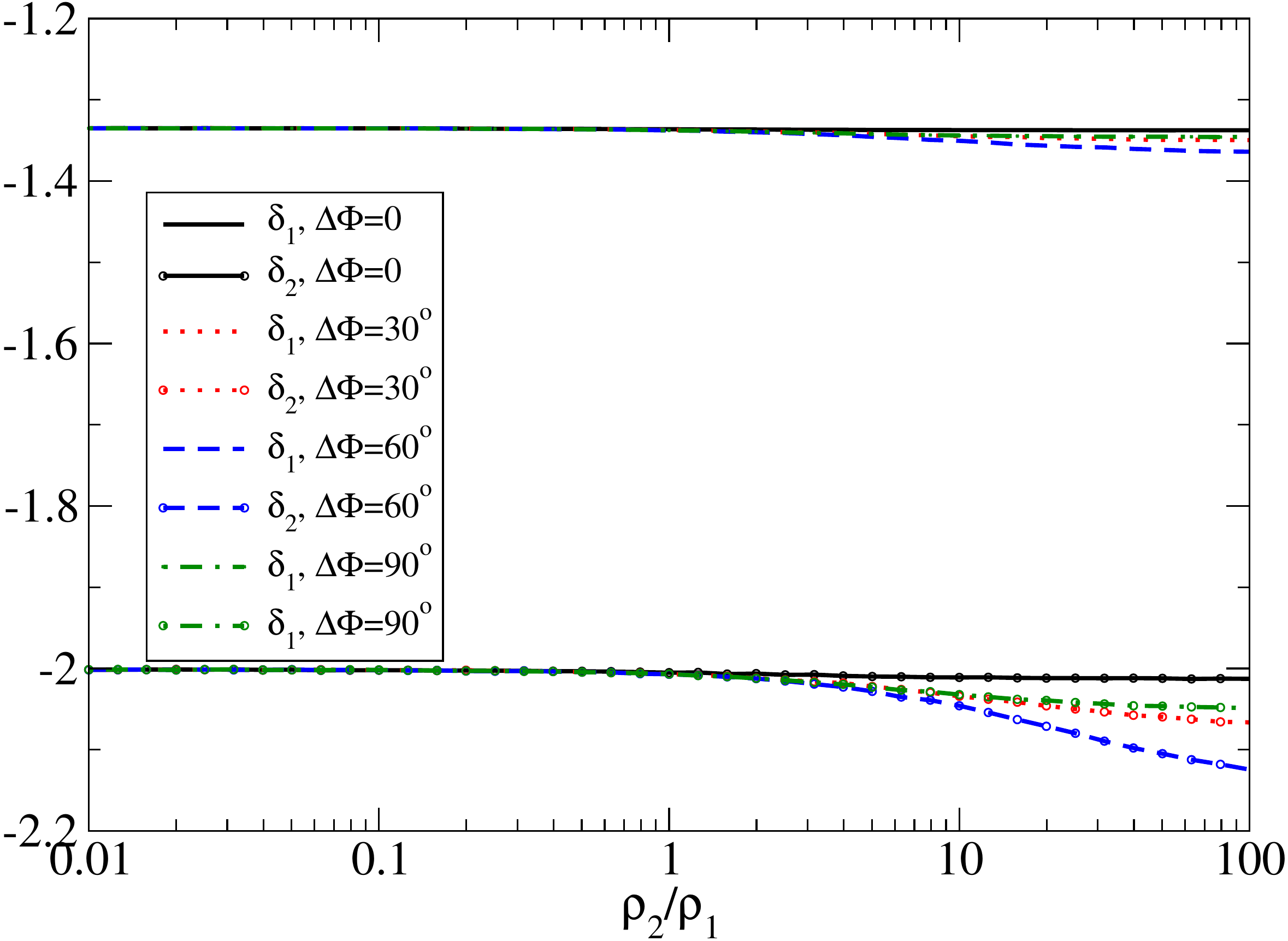} 
  \end{array}$
\caption{For several values of the density ratio $\rho _2/\rho _1$ and angular separation between 
the major axes of the two bars, $\Delta \Phi$, the variation of three representative parameters of 
bar 1 under the gravitational influence of bar 2 are plotted. The panels show:  
$\Omega _1/\sqrt{\pi G\rho _1}$ (top) ;  $q_{21}$, $q_{31}$ (middle);  $\delta _1$ (bottom) on top of 
the panel, and $\delta _2$ (see text for their definition). The two bars are set to be of equal axis 
lengths ($q_{2i}=0.520, q_{3i}=0.394$) and so would have the same rotational velocity
 ($\Omega _i^2/\pi G\rho _i=0.394$) in the Jacobian case.}
\label{Fig:modJac}
\end{figure}

This approach is somewhat simplistic but is sufficient for our purposes as we just try to discuss 
basic properties of the system of two bars. A global discussion of the interaction must also 
consider more realistic non-homogeneous cases and contain other elements, such as the interaction 
with other galactic components (disc, halo, etc.), and address the question of the deviations from 
equilibrium. The equilibrium and stability in self-gravitating systems such as bars in spiral 
galaxies (with violent relaxation, persistence of metaequilibrium states, slow collisional dynamics,
phase transitions, etc.) has also been explored by other authors, getting in some cases only small 
departures of the system from equilibrium in an oscillation as a radial pulsation  
\citep{Christodoulou1995, Chavanis2005}. The relaxation time to reach the quasi-equilibrium 
states in the interaction of two bars is also interesting to explore, while the concept
 of relaxation is somewhat difficult to apply in our simple model, as we do not have orbital 
support in the bars, so that the internal kinematics of the system is simply not considered. In any case, 
double bar systems in galaxies are observed, and these may be regarded, at least to a first 
approximation, as ellipsoids in  equilibrium.

When the bars are not rigid, the governing equation is Eq.\ (\ref{eq_ppal}), but in this case the 
moments of inertia of each bar, $I_i$, vary following Eq.\ (\ref{eq_ppal}) with the changes in the 
ellipsoid geometrical parameters, $(a_1\ a_2\ a_3)$. That is to say, the interaction between the 
two systems implies the modification of the velocity patterns due to the interaction; also, 
both systems accommodate their own mass distribution to the new conditions to reach a new 
equilibrium state, which we have prescribed to be a triaxial ellipsoid. Taking this into account, 
each rotational velocity in Eq.\ (\ref{eq_ini}) transforms into the following form:

\begin{equation}
 \dot \Omega _i = \tau _i (\Delta \Phi )\left(I_i + \frac{\mathrm{d}I_i }{\mathrm{d}\Omega_i }\Omega_i\right)^{-1}
\label{eq_ppal_def}
\end{equation}

For ellipsoids, the moments of inertia are given by Eq.\ (\ref{eq_moi}). Hence,

\begin{equation}
\dot \Omega_i = \frac{\tau_i (\Delta \Phi )}{I_i}\left(1+\frac{2\Omega _i}{a_{1i}}\frac{\partial 
a_{1i}}{\partial \Omega _i}+\frac{2q_{2i} \Omega_i }{1 + q_{2i}^2 }\frac{\partial {\kern 1pt} q_{2i} }{\partial {\kern 1pt} \Omega_i } \right)^{-1}
\label{eq_ppal_defj}
\end{equation}

The gravitational potential for each of the bars can be obtained by adding the potential created by 
its own mass, ${\cal V}_i$, i.e.\ the potential of a homogeneous ellipsoid at points interior to itself, 
and the one due to the other bar, ${\cal V}_{j \ne i}$, i.e. the potential of an homogenous 
an ellipsoid at points interior or exterior to itself. Each of these follows the same equation, with 
different integration limits, which turn out to be:

\begin{equation}
\begin{array} {l}
{\cal V}(\mathbf{x})=-\frac{3}{4}GM\int _{\nu(\mathbf{x})}^\infty\left[1-\sum_{j=1}^{3}\frac{x_j^2}{a_j^2+\lambda}\right] \frac{d\lambda }{\Delta(\lambda)},\\
 \vspace*{-0.3cm} \\
\Delta^2 (\lambda)=(a_1^2+\lambda)(a_2^2+\lambda)(a_3^2+\lambda)\\
\end{array}
\label{pots}
\end{equation}
$M$ and $a_j$ being the mass and semi-axes of the bar generating the potential, $x_j$ the 
$xyz$ coordinates of the position at which the potential is being evaluated ($\mathbf{x}$), and 
$\nu(\mathbf{x})$ is zero for points at the surface and interior to the bar and is the 
ellipsoidal coordinate of $\mathbf{x}$ for exterior points, i.e.\ the positive root of

\begin{equation}
\sum_{j=1}^{3}\frac{x_j^2}{a_j^2+\nu}=1
\label{ellcor}
\end{equation}

This potential can be expressed in a more convenient form, with the change $\lambda=a_{1i}^2\ \mu$, as

\begin{equation}
\begin{array} {l}
{\cal V}_i(\mathbf{x})=\frac{3}{4}\frac{GM_i}{a_{1i}^3} \times \left[A_i[u_i(\mathbf{x})]x^2+B_i[u_i(\mathbf{x})]y^2+\right.\\
 \vspace*{-0.3cm} \\
\left.\hspace*{1.25cm}+C_i[u_i(\mathbf{x})]z^2 -D_i[u_i(\mathbf{x})]a_{1i}^2\right] \\
 \vspace*{-0.3cm} \\
A_i[u_i(\mathbf{x})]=\int _{u_i(\mathbf{x})}^\infty \frac{d\mu }{(1+\mu )\Delta' _i(\mu)} \\
 \vspace*{-0.3cm} \\
B_i[u_i(\mathbf{x})]=\int _{u_i(\mathbf{x})}^\infty \frac{d\mu }{(q_{2i}^2+\mu )\Delta' _i(\mu)} \\
 \vspace*{-0.3cm} \\
C_i[u_i(\mathbf{x})]=\int _{u_i(\mathbf{x})}^\infty \frac{d\mu }{(q_{3i}^2+\mu )\Delta' _i(\mu)} \\
 \vspace*{-0.3cm} \\
D_i[u_i(\mathbf{x})]=\int _{u_i(\mathbf{x})}^\infty \frac{d\mu }{\Delta' _i(\mu)} \\
\vspace*{-0.3cm} \\
u_i(\mathbf{x})=\nu_i(\mathbf{x})/a_{1i}^2 \ ;\ \Delta' _i(\mu)^2\\
\hspace*{.9cm}=(1+\mu)(q_{2i}^2+\mu )(q_{3i}^2+\mu )
\end{array}
\label{potencial}
\end{equation}
where the subindex $i$ indicates that the magnitudes are evaluated using the parameters of bar $i$. 
Note that $A_i$, $B_i$, $C_i$ and $D_i$ are dimensionless, and that they do not depend on $a_{1i}$ 
for the points interior to the bar. $D_i$ cancels out in Eq.\ (\ref{dyn2def}) as it does not depend 
on the spatial coordinates. In Fig.\ \ref{Fig:potbar} there are examples of the shape of this 
potential evaluated in the XY plane of the bar and at several azimuths measured from the bar's
 principal axis. It can be observed how this grows more or less linearly with radial distance
 from the centre of the ellipsoid until the surface of the body is reached ($R/a_1=1$), from which point
it fades with distance, as expected.

Finally, we apply the condition for hydrostatic equilibrium of each of the rotating bar, 

\begin{equation}
{\cal V}_i(\mathbf{x_i})+{\cal V}_j(\mathbf{x_i})-\frac{1}{2}\Omega _i^2(x_i^2+y_i^2)={\rm constant}
\label{equilibrio}
,\end{equation}
with $j\ne i$, which are combined with the requirement of the bars being triaxial ellipsoids 
to derive the governing equations of equilibrium of the rotating bar under the influence
 of its own gravity and that of the other bar. Hence Eq.\ (\ref{equilibrio}) is developed for 
the points of the surface of each bar along the three axes, and with a second external bar 
separated by an angle $\Delta \Phi $, to get:

\begin{equation}
\begin{array}{l}
A_i[0]+MR_{ji}\left[A_{j}[u_{ij1}]\cos ^2(\Delta \Phi)+\right. \\
\left.+B_{j}[u_{ij1}]\sin ^2(\Delta \Phi )\right] -E_i=\\
=q_{2i}^2\left\{B_i[0]+MR_{ji}\left[A_{j}[u_{ij2}]\sin ^2(\Delta \Phi)\right.\right. \\
\left.+B_{j}[u_{ij2}]\cos ^2(\Delta \Phi )\right]-E_i\big\}\\
=q_{3i}^2\left\{C_i[0]+MR_{ji}C_{j}[u_{ij3}]\right\} \\
MR_{ji}=\frac{M_j}{M_i}f_{ij}^3
,\ 
E_i=\frac{2\Omega _i^2a_{1i}^3}{3GM_i}
,\ 
f_{ij}=\frac{a_{1i}}{a_{1j}}
\end{array}
\label{dyn2def}
\end{equation}
where, again, $j\ne i$. The $u$ parameters in Eq.\ (\ref{dyn2def}) are given by Eq.\ (\ref{ellcor}), 
followed by $u_i(\mathbf{x})=\nu_i(\mathbf{x})/a_{1i}^2$,  applied in sequence to the three end 
points of the principal axes of bar $i$,  using the parameters of bar $j$, to obtain:
\[\begin{array}{lll}
u_{ij1}&=&\max \left\{0,\frac{1}{2}f_{ij}^2-\frac{1+q_{2j}^2}{2}+\sqrt{\Delta _{ij1}}\right\} \\
\Delta _{ij1}&=&\left[\frac{1+q_{2j}^2}{2}-\frac{1}{2}f_{ij}^2\right]^2-q_{2j}^2+f_{ij}^2\left[\sin ^2(\Delta \Phi )+\right. \\
&&\left.+q_{2j}^2\cos ^2(\Delta \Phi)\right] \\
u_{ij2}&=&\max \left\{0,\frac{1}{2}f_{ij}^2q_{2i}^2-\frac{1+q_{2j}^2}{2}+\sqrt{\Delta _{ij2}}\right\} \\
\Delta _{ij2}&=&\left[\frac{1+q_{2j}^2}{2}-\frac{1}{2}f_{ij}^2q_{2i}^2\right]^2-q_{2j}^2+f_{ij}^2q_{2i}^2\times \\
&&\times\left[\cos ^2(\Delta \Phi )+q_{2j}^2\sin ^2(\Delta \Phi)\right] \\
u_{ij3}&=&\max \left\{0,f_{ij}^2q_{3i}^2-q_{3j}^2\right\}
\end{array}\]

In Eq.\ (\ref{dyn2def}) it becomes clear that the influence of bar $j$ over bar $i$ is governed 
by the ratio of masses given by $MR_{ji}$. If set to 0, it is restricted to the case of Jacobi 
ellipsoids, and we get:

\begin{equation}
\Omega _i^J=\sqrt{\frac{3GM_i}{2a_{1i}^3}\frac{A_i[0]-q_{2i}^2B_i[0]}{1-q_{2i}^2}}
\label{omega_jacobi}
\end{equation}
which corresponds to the rotational velocity of a Jacobi ellipsoid. By definition, 
$\partial A_i[0]/ \partial a_{1i}=\partial B_i[0]/ \partial a_{1i}=0$, and the quantity 
$$\left[\frac{A_i[0]-q_{2i}^2B_i[0]}{1-q_{2i}^2}\right]=\int _{0}^\infty \frac{d\mu }{(1+\mu )(q_{2i}^2+\mu)\Delta _i(\mu)}$$ 
decreases when $q_{2i}$ increases. Hence, $\Omega _i^J\times \partial q_{2i}/\partial \Omega_i^J<0$, 
which means that the faster the rotation, the higher the length is  and, in consequence, the 
eccentricity of the ellipsoids. Using Eq.\ (\ref{omega_jacobi}),
 $(2\Omega _i^J/a_{1i})\times(\partial a_{1i}/\partial \Omega _i^J)= -4/3$. Therefore, looking
 at Eq.\ (\ref{eq_ppal_defj}), we see that $\dot\Omega_i^J/\tau_i <0$. The torque exerted by the
 bars on each other, in the case of pure Jacobi ellipsoids, produces a change in velocity of opposite 
direction to that of the torque. This means that the bars seem to repel each other, contrary 
to what happens in the rigid case, where the rotational velocity follows the course of the torque. 
Clearly, this is not an accurate model of the double bar configuration, as the potential of the 
other bar is not being considered in the equilibrium figure of the bar under study.

It is instructive to explore whether in the general case given by Eq.\ (\ref{dyn2def}), which can 
be denoted as {\it modified Jacobi ellipsoids} following \citet{Miyamoto1967}, the same qualitative 
behaviour can be observed, as in the case of pure Jacobi ellipsoids. Figure \ref{Fig:modJac} 
illustrates the scenario for a particular case with a given set of geometrical and kinematic 
parameters. It can be seen that $\delta _1\equiv (2\Omega _i/a_{1i})\times(\partial a_{1i}/\partial \Omega _i)<-1$ 
and, as a matter of fact, very close to $-$4/3; and $\delta _2\equiv \Omega _i\times(\partial q_{2i}/\partial \Omega_i)<0$ 
as well, which implies from Eq.\ (\ref{eq_ppal_defj}) that $\dot\Omega_i/\tau_i<0$ as in the
 Jacobian case. In Fig.\ \ref{Fig:modJac} the ratio of bar densities, which governs the influence 
of the external bar over the bar under study via the parameter $MR_{ji}$ in Eq.\ (\ref{dyn2def}),
 runs up to 100. So even for a very strong influence of bar 2 on bar 1 this result is maintained, 
Which means that the same behaviour as seen in the Jacobian case is qualitatively followed here: 
the bars seem to repel each other, and this result is not dependent on the axial values $a_{ki}$.

We also see in the middle panel of Fig.\ \ref{Fig:modJac} that $q_{21}$ is not very much
 affected with respect to the nominal value for the Jacobian case, the variation being roughly 
10\% at most, whereas the effect in $q_{31}$ is more conspicuous, reaching about 30\%. But 
it is the rotation velocity, $\Omega _1$, which is affected the most, in terms of departure 
from the value at equilibrium of a Jacobi ellipsoid, which is located on the left of each 
of the three panels. This is particularly true for $\Delta \Phi \approx 0$ and large secondary bar masses, 
and reflects how that bar significantly increases the total gravitational attraction 
towards the centre of the galaxy, which must be compensated by a larger rotation velocity.

The evolution with time can be followed using Eqs (\ref{eq_ppal_defj}) and (\ref{dyn2def}), 
and the conservation of the total angular momentum. In this paper we have taken the approximation
 that bar 2 is much more massive than bar 1, so that their mutual influence is mostly reflected in
 bar 1. In this model, bar 2 can always be considered as a Jacobi ellipsoid with a well prescribed 
relationship between the geometry, i.e.\ the axes, and the rotational velocity, whereas bar 1 is a 
modified Jacobi ellipsoid described by Equation (\ref{dyn2def}). We have also made the additional 
simplification of considering bar 2 as having fixed axes, which can be justified by the small impact of
 bar 1 on its geometry, as can be seen in Fig. \ref{Fig:modJac}/middle for small values of the
 density ratio. This small variation is even neglected for the ease of the calculationi. Finally,
$\delta _1$ has been prescribed to take a fixed value of $-$4/3, which is roughly what we see in 
Fig.\ \ref{Fig:modJac}/bottom over an ample range of density ratios. Hence, for the sake of simplicity, 
bar 2 will be treated as a rigid bar described in section \ref{.srig}.

As initial parameters for the multistep integration method to proceed, we have selected the following:
\begin{itemize}
\item The masses, $M_1$ and $M_2$, of the intervening bars, which remain unaltered in the integration.
\item The length of the axes, ($a_1,a_2,a_3)$. Using Eqs \ref{omega_jacobi}, for bar 1, or 
\ref{dyn2def}, for bar 2, the initial rotational velocity can be determined.
\item The initial angle, $\Delta \Phi$, between the major axes of the two bars.
\end{itemize}

Now the integration can proceed in steps of fixed amount of time, as mentioned before. 
The first task is to calculate the torque between the bars, as described in Appendix
 \ref{app_torq}. The torque between the two intervening bars is independent of the degree 
of deformability of the system, as it reflects the action between the two systems in a given 
structural configuration, since it is also independent of the relative speeds between bars. It is 
only dependent on the angle between bars and the given mass distribution. So the prescriptions 
in Appendix \ref{app_torq} provide the recipe for the torque, whose run will follow the pattern 
as in Figure \ref{fig_torq}. Next, Eq.\ (\ref{eq_ppal_defj}), on which the derivatives are numerically
  calculated, provides the new value for the rotational velocity of bar 1, whereas the corresponding 
value for bar 2 comes directly from the torque (see Eq.\ (\ref{eq_ini}). Finally, Eq.\ (\ref{dyn2def}) 
gives the new axes for bar 1, while those of bar 2 are fixed. The whole process is repeated for a 
number of time steps to derive the evolution of the double bar system.

\begin{figure}[ht]
  $ \begin{array}{l}
  \includegraphics[angle=0,width=0.47\textwidth]{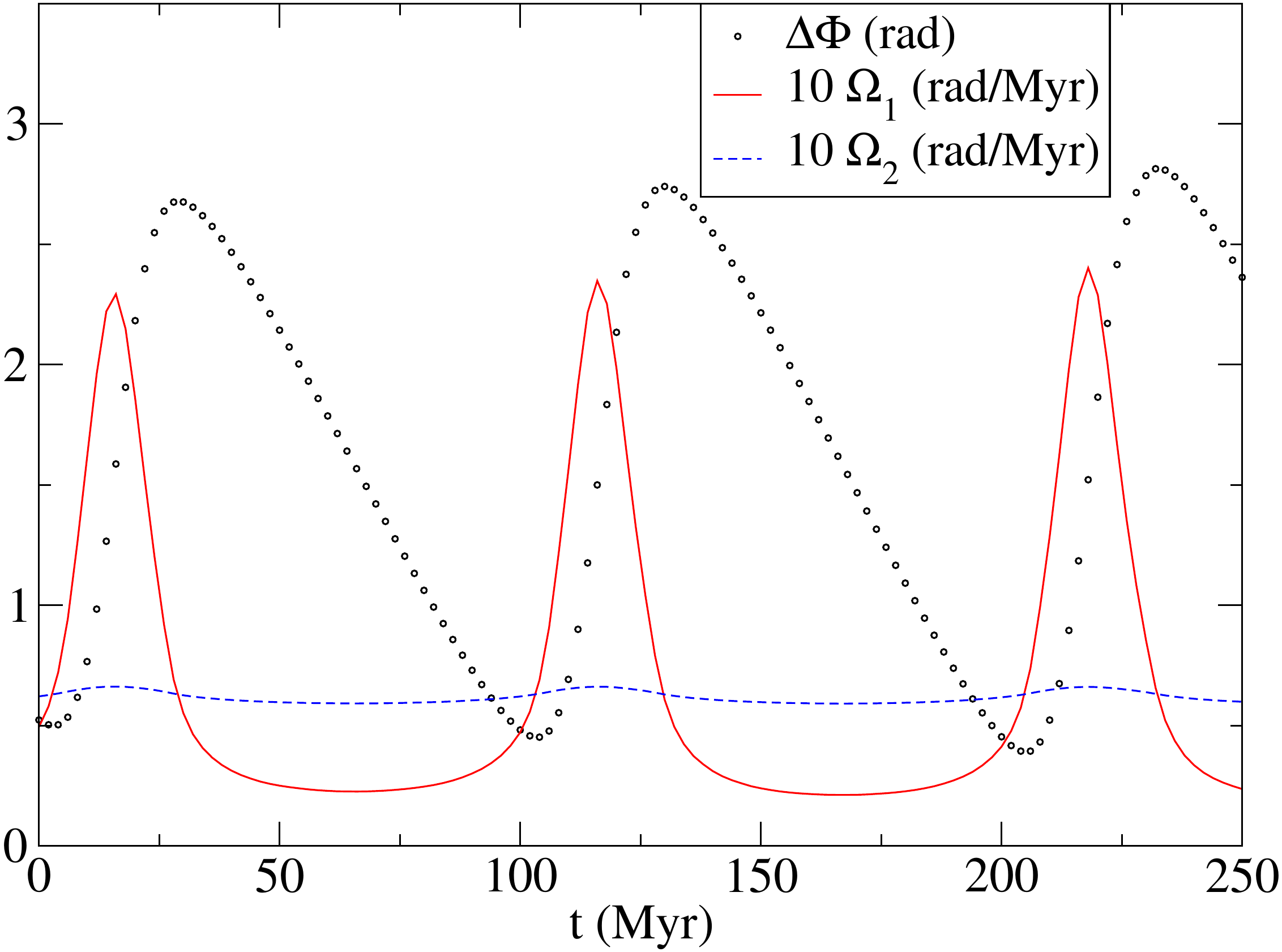}  \\
   \includegraphics[angle=0,width=0.47\textwidth]{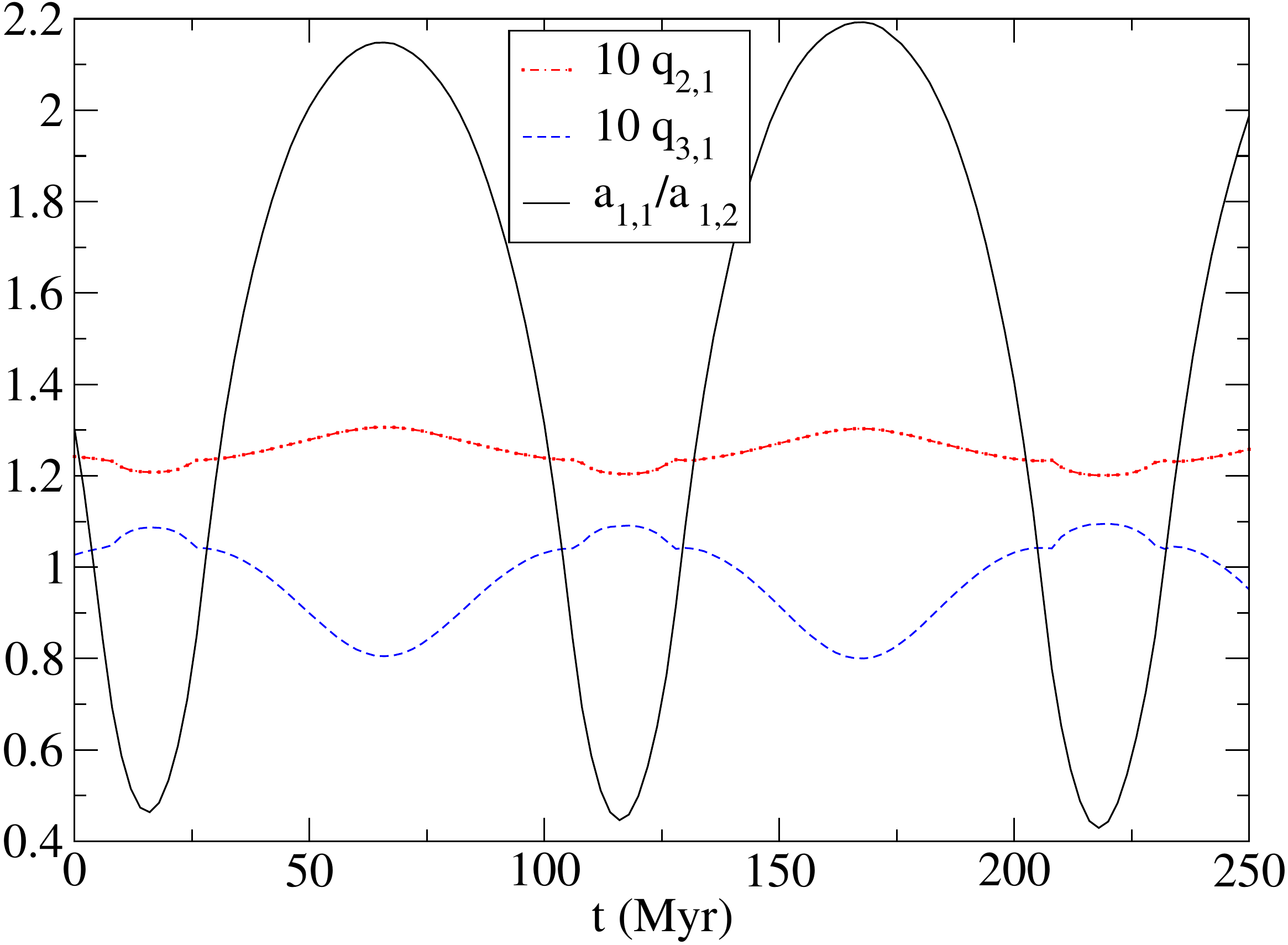}
   \end{array}$
\caption{Evolution of two bars with $M_1=3\times 10^9, M_2=2\times 10^{10}$ solar masses and initial
parameters for bar 1: $a_{11}=4$ kpc, $q_{21}=0.120, q_{31}=0.115$; and a fixed geometry for bar 2 of 
$a_{12}=3$ kpc, $q_{22}=0.520, q_{32}=0.394$, with initial velocity of $\Omega _2^2/\pi G\rho _2=0.316$ 
(Jacobi ellipsoid); $\Delta \Phi =30^\circ$ as initial value. Top: evolution of the angular separation, 
in the range [0, $\pi$] and the angular velocities. Bottom: evolution of the axial ratios and semimajor 
axis of bar 1.}
\label{Fig:evol1}
\end{figure}

In Fig.\ \ref{Fig:evol1}, we see the evolution of bar 1 for a case with initial morphology 
$a_{11}=4, q_{21}=0.12, q_{31}=0.115$; and bar 2 with a fixed values of 
$a_{12}=3, q_{22}=0.520, q_{32}=0.394$ (lengths in kpc). The initial separation is 
$\Delta \Phi =30^\circ $, and masses $M_1$ and $M_2$  equal to $3\times 10^{9}$ M$_\odot $ and $2\times 10^{10}$ 
M$_\odot $ respectively. The time step for the integration has been set to 0.5 Myr. In this case the density
 ratio is initially 0.94, from which no conclusions can be derived by looking at 
Fig.\ \ref{Fig:modJac}, as the geometry of the bars are quite different. It is the mass ratio, 
which remains constant at 6.67 (the figure for $MR_{21}$ in Eq.\ (\ref{dyn2def})), which sets 
the validity of the simplifying assumptions explained before. In Fig.\ \ref{Fig:evol1} it is
seen how at first bar 1 decreases its size and increases its angular velocity as a result
of the initial repulsive torque produced by bar 2, which quickly becomes $\Omega _1>\Omega _2$.
 This decrease in size of bar 1 is approximately constant in aspect ratio, as the major axis 
decreases by a factor greater than two while the $q$-axial ratios experienced minor changes. It 
can also be seen how the angular separation initially becomes smaller than the initial setting 
owing to $\Omega _2$ being larger than $\Omega _1$, a situation that is rapidly inverted by the 
action between the bars.

Soon after this episode of minimal size of bar 1, which corresponds to the maximum of rotational 
velocity, bar 1 quickly starts to approach the opposite end of bar 2, owing to its high relative
 velocity, and begins to experience a torque that tends to reverse the motion of bar 1 with respect 
to bar 2. $\Omega _1$ decreases rapidly until it becomes smaller than $\Omega _2$, when the bars 
start to invert their relative movement. This decrease in $\Omega _1$ is accompanied by an increase 
in $a_{11}$, which again happens with approximately constant aspect ratio. The result is a periodic
oscillation between the states of minimum size--maximum angular velocity and maximum size--minimum 
angular velocity.

\begin{figure}[ht]
  $ \begin{array}{l}
  \includegraphics[angle=0,width=0.47\textwidth]{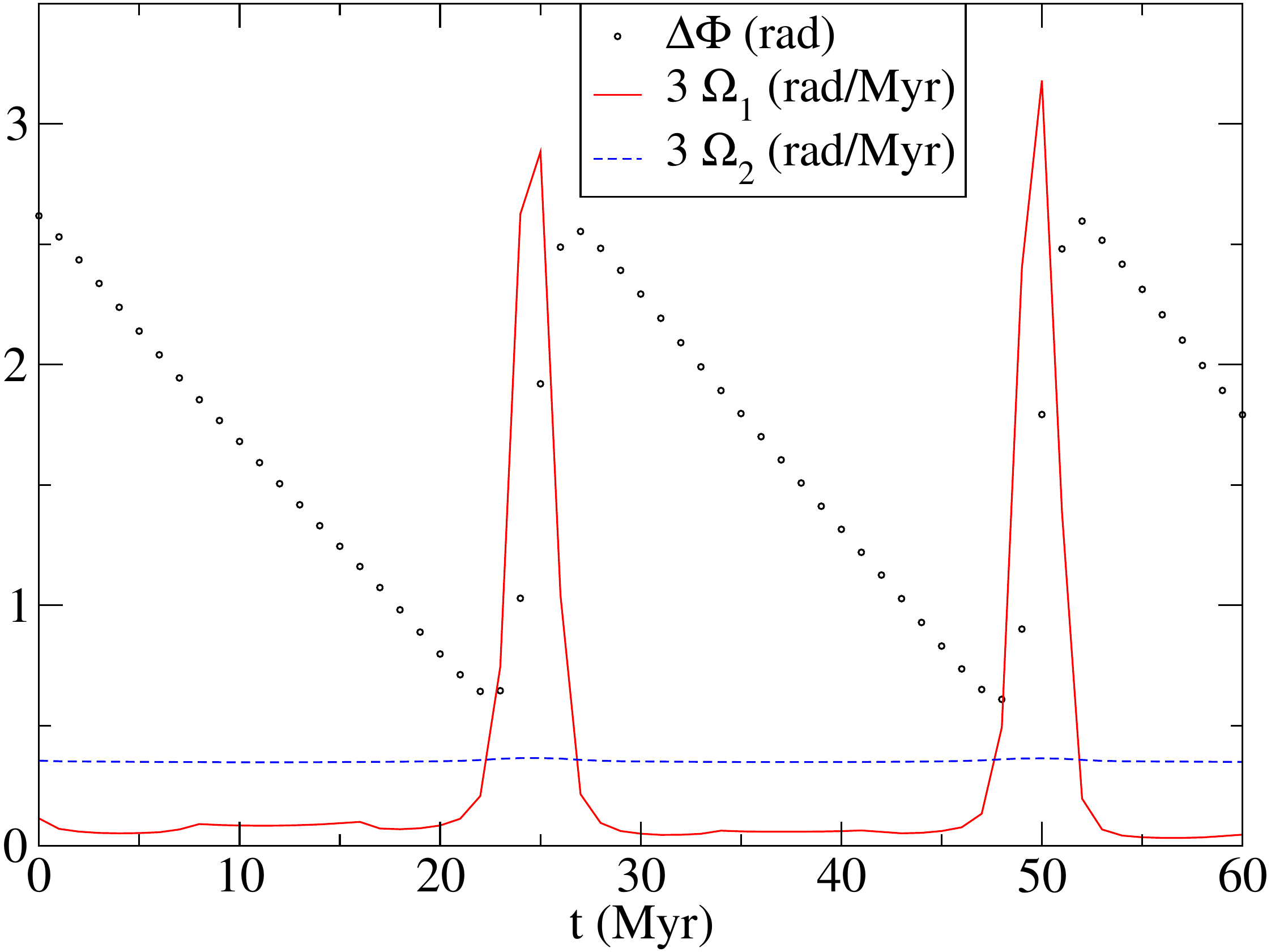} \\
   \includegraphics[angle=0,width=0.47\textwidth]{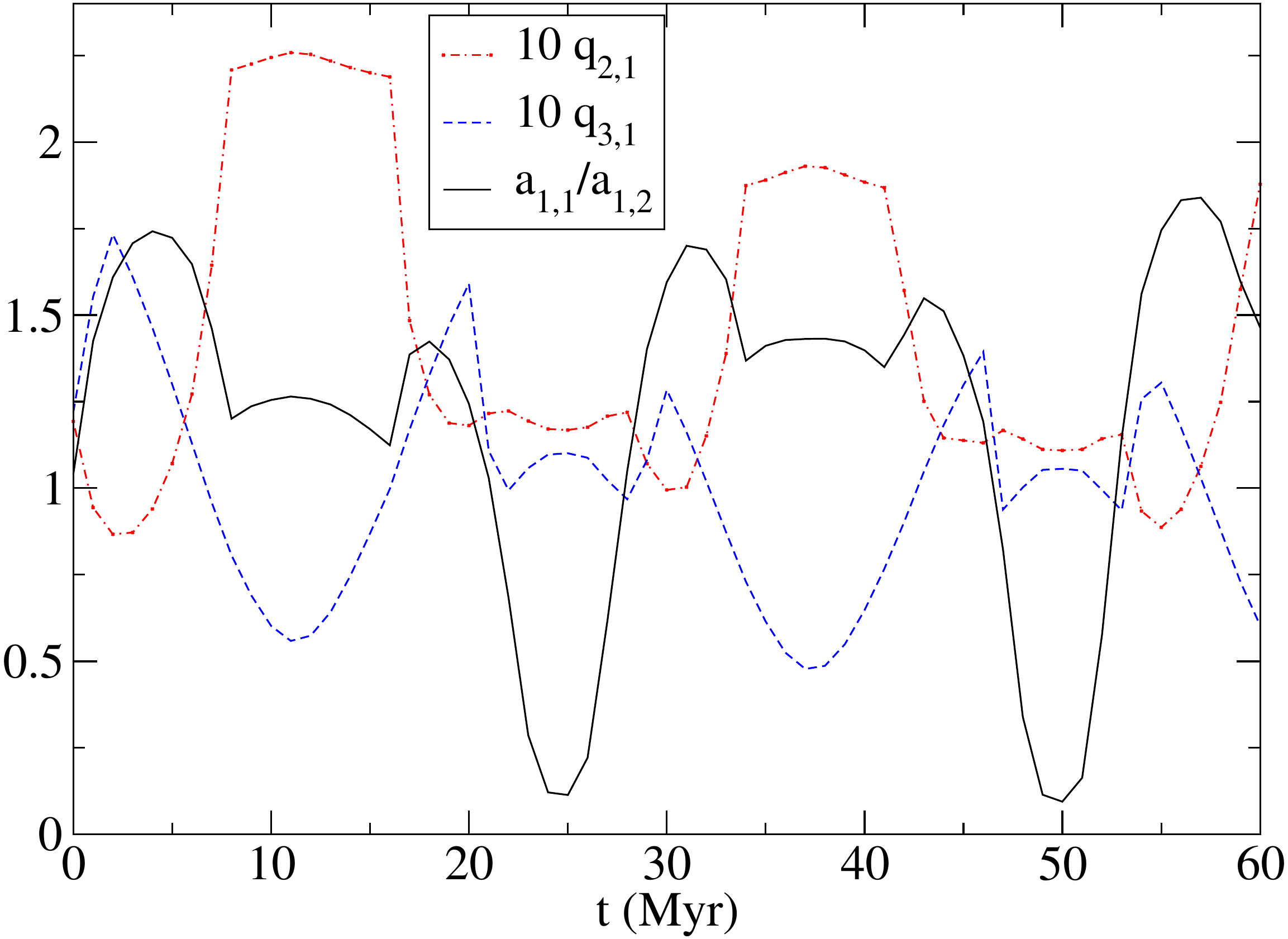}  
  \end{array}$
\caption{
Evolution of two bars with $M_1=3\times 10^9, M_2=2\times 10^{10}$ solar masses, and initial
 parameters: $a_{11}=3$ kpc, $q_{21}=0.120, q_{31}=0.115$; and a fixed geometry for bar 2 
equals to that of the initial bar 1, which implies an initial velocity of $\Omega _2^2/\pi G\rho _2=0.0773$ 
(Jacobi ellipsoid); $\Delta \Phi =-30^\circ$ as initial value. Panels as in Figure \ref{Fig:evol1}. }
\label{Fig:evol2}
\end{figure}

In Fig.\ \ref{Fig:evol2} another configuration of a double bar system is explored, now with the two 
bars initially sharing  the same geometry, which remains fixed for bar 2. At the beginning, the 
morphology of bar 1 is given by $a_{11}=3$ kpc, $q_{21}=0.12, q_{31}=0.115; a_{12}=3$. The initial 
separation is $\Delta \Phi =-30^\circ $, and masses $M_1=3\times 10^{9}$ M$_\odot $ and 
$M_2=2\times 10^{10}$ M$_\odot $, as in the previous case. The time step for the integration has
been set to 0.1 Myr, as this configuration shows higher complexity. In this case of equal geometry 
the initial density ratio of 0.15 permits the use of Fig.\ \ref{Fig:modJac} to ascertain the relative
influence of each opposite bar.

At the beginning, the size of bar 1 grows and its aspect changes quite strongly: $q_{21}$ becomes 
smaller whereas $q_{31}$ increases, which makes the vertical axis larger than the second axis. The 
rotation during this episode slows down and comes close to stopping due to the positive torque. The angular
 separation of the two bars increases as the faster rotation of bar 2 tends to get bar 1 closer to the
other end. As this is happening, the torque becomes smaller because of the larger separation and bar 1 
becomes reduced in size, recovering angular velocity. As before, the change in size is accompanied by 
a substantial variation in aspect, as $q_{21}$ and $q_{31}$ vary in an opposite sense by a factor greater than 
two. As the bars continue approaching, the torque increases and the rotation of bar 1 also increases
 accordingly until the bars reach their minimum separation in which the larger torque pushes bar 1 to a much 
faster rotation, over that of bar 2, which recovers the original location of the bars. The system then 
develops in a periodic oscillation as in the case of Figure \ref{Fig:evol1}. The case appears more complex
 because of the variation of the axial ratios is here more important, taking a large factor of the angular
 momentum interchange. The shape of the global system is not exactly the same in each period of the 
oscillation in Fig.\ \ref{Fig:evol2}, but this is most probably due to inaccuracies in the numerical 
calculations, which we will improve in future papers. As far as we can see, extending the evolution of
the system to a much larger time period than shown in Fig.\ \ref{Fig:evol2}, the system is bounded between
the two status of maximum and minimum angular separation, while the exact details of the interaction would 
necessitate also accounting for the deformations of bar 2.

The two cases depicted in Figs \ref{Fig:evol1} and \ref{Fig:evol2} both share the same feature in the 
motion pattern of bar 1, the lightest of the two bars and the live one in the simulations. Since the 
two cases correspond to configurations in which bar 1 remains trapped with an oscillatory motion around 
bar 2 ($\Delta \Phi$ is confined in a given range) and always revolving in the same sense ($\Omega_1$ 
does not change its sign), two different events of alignment and perpendicularity can be distinguished. 
Initially, bar 2 gets an impulse with different sign with respect to its rotational velocity in the 
two examples. While for the case depicted in Fig.\ \ref{Fig:evol1} the repulsion between the bars pushes 
to increase the frequency; in the case of Fig.\ \ref{Fig:evol2} the torque slows down the motion of bar 2.
 This is due solely to the different initial location of the two bars in the examples. The change in 
angular velocity is accompanied by a change in opposite sense in bar size ($a_{11}$), as dictated by 
Equation (\ref{eq_ppal_def}). In the first example, bar 1 keeps increasing the speed and decreasing the size,
moving until the closest approach on the other side and both bars become perpendicular to each other, 
when the mutual torque inverts its sign and bar 1 starts to brake its rotation and increases its size 
accordingly. Bar 1 becomes slower than bar 2 and the two bars tend towards separation. The perpendicular 
alignment in this motion marks the starts of the acceleration and size decrease of bar 1. The pattern 
is then repeated. So the status of the living bar in the positions close to being in alignment and orthogonal 
alternates during the motion pattern. Perpendicularity corresponds alternately to maximum (minimum) 
and minimum (maximum) rotation speed (size). In the second example, Fig.\ \ref{Fig:evol2}, the situation
is similar but, owing to the different initial relative position, the evolution starts reaching first the
status of minimum speed, and maximum size, of the living bar when the two bars are perpendicular to each other.

\begin{figure}[ht]
  $ \begin{array}{l}
  \includegraphics[angle=0,width=0.47\textwidth]{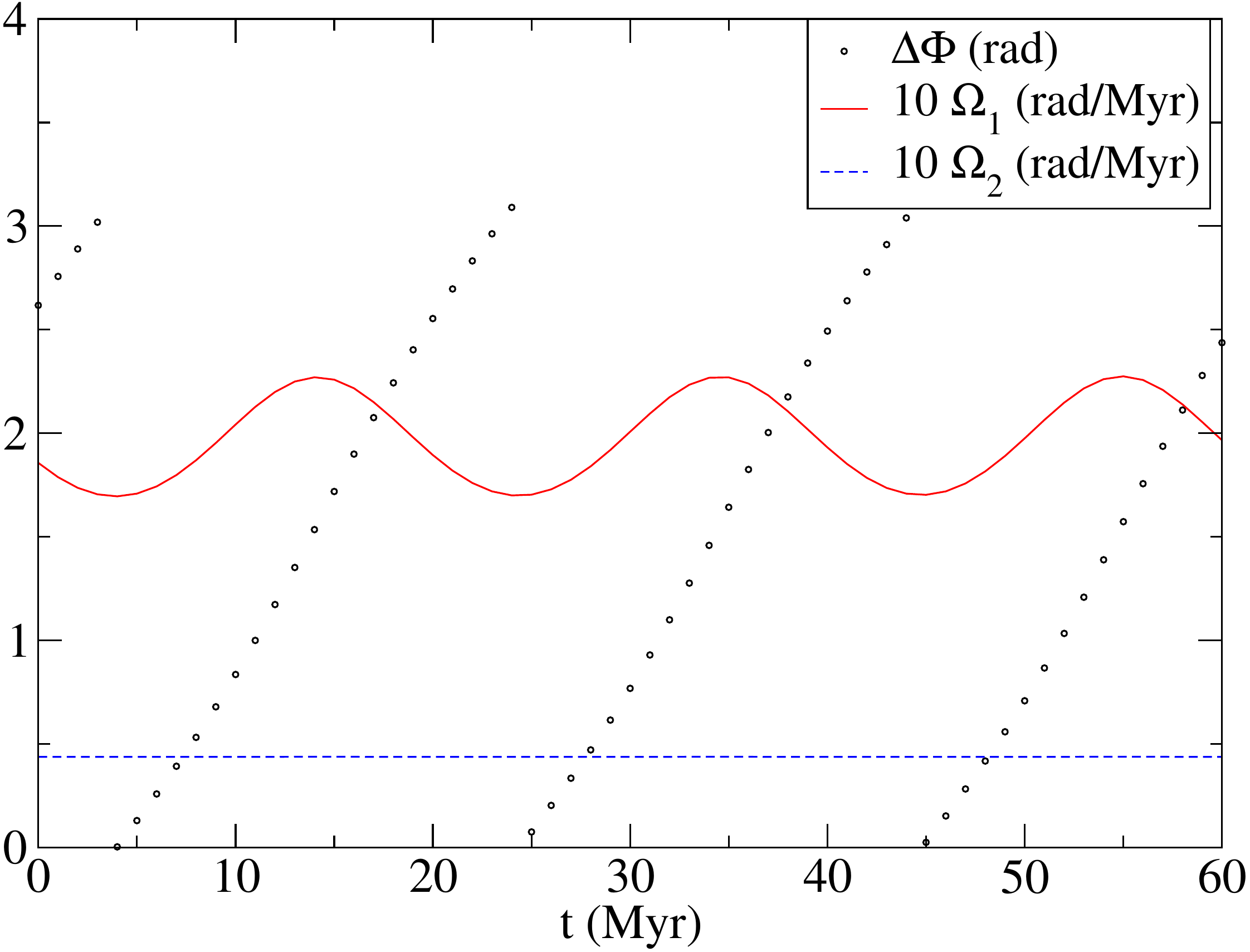} \\
  \includegraphics[angle=0,width=0.47\textwidth]{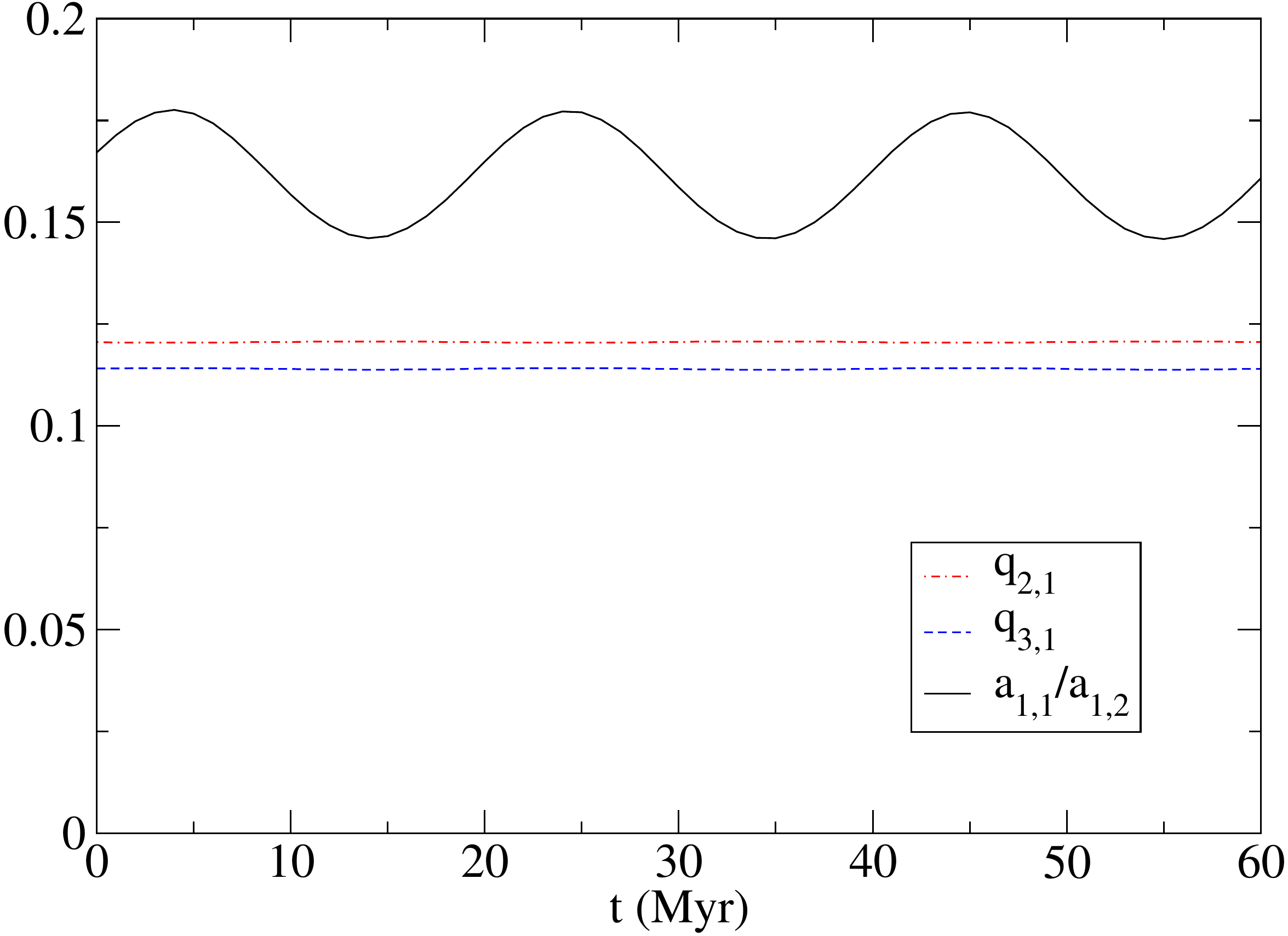} 
  \end{array}$
\caption{
Evolution of two bars with $M_1=2\times 10^8, M_2=10^{10}$ solar masses, and initial parameters:
 $a_{11}=0.5$ kpc, $q_{21}=0.12, q_{31}=0.115$; bar 2 remains unaltered with $a_{12}=3$ kpc,
$q_{22}=0.52, q_{32}=0.393$; $\Delta \Phi =-30^\circ$ as initial value. Panels as in 
Figure \ref{Fig:evol1}. This configuration might represent a secondary bar inside the central bulge as in DS07.}
\label{Fig:evol3}
\end{figure}

To compare the predictions of the approach presented in this paper against $N$-body simulations, we 
have prepared a third configuration following the figures of DS07. This simulation, whose results 
are shown in Fig.\ \ref{Fig:evol3}, attempts to model a small secondary bar (bar 1) embedded within
 a larger and more massive primary bar (bar 2). In this case, bar 1 is always at the interior of
bar 2, due to its small size, hence both potentials in Eq.\ (\ref{equilibrio}) are calculated at 
interior points. As seen in Fig.\ \ref{Fig:evol3}, the result is that bar 1 develops an oscillatory
 motion pattern, with period of 20\ Myr, much in the same way as fig.\ 2 of DS07, once B1 and B2
have been set. In this case, $\Omega_1$ is always faster than $\Omega_2$, as in DS07,  so bar 1 is
revolving around bar 2 and there is no alternating of maximum and minimum frequency between the 
positions of perpendicularity or alignment. The peak of rotation speeds of bar 1 always coincides 
with the bars being perpendicular and the minimum value, with the bars in alignment. This is 
contrary of what happens in the runs of DS07, where the small bar gets its maximum speed when the 
two bars align, which also coincides with the minima of the Fourier $m=2$ amplitude. This contradictory
result can be attributed to the prescriptions of the modified Jacobi ellipsoid model, which acts as if
 the torque between the two bodies were of  reversed sign. So the opposite results between DS07 and 
this paper in the locus of the maxima/minima of the rotation speeds was in a way expected. Except for 
this, the trend of the motion of the smaller bar is quite similar in both models, considering the
 rotation period and the amplitude of the oscillations. Further commonalities cannot be expected in 
view of the large difference in the conditions of the two scenarios.

As mentioned before, our approach is rather simple and lacks important ingredients to mimic
 the real world, such as the galactic disc and the halo, and as such does not aim to
to obtain the exact behaviour of the bars. Instead, the analytic approach that we have followed 
permits us to keep track of the evolution of the system in all its phases and gain understanding of the
 phenomena governing its evolution. Precisely because of that, an analytical approach, even 
an approximate one, is useful, and obtaining different results with respect to $N$-body simulations does 
not imply that its results are necessarily wrong. On the contrary, it may help in pointing out
some facts that may be difficult to catch in a numerical simulation.

We have checked with other parameter sets and get general results of the same type as
those described previously. A long bar varies its morphology and velocity pattern under the influence
of a more massive and thicker bulge: oscillating between a shorter bar nearly embedded in the bulge, 
and a longer bar with very small angular velocity.

\section{Conclusions}
We have modelled the behaviour of two isolated bar-like mass distributions under the effect of
their mutual gravitational attraction. The rigid case produces coupling of the two bars over a 
rather wide range of mass ratios and angular velocities. The whole system will continue rotating 
in an ordered way interchanging angular momenta during the coupling. Only in cases with large 
differences in rotational velocities will coupling not occur.

The case of deformable systems has been considered with the assumption of the intervening bars being 
homogeneous ellipsoids following a fixed relationship in their geometry derived from the 
equilibrium of gravitational interactions from the two bars and the centrifugal forces (modified
Jacobi ellipsoids), thus determining their angular speed. This case has been modelled under 
the simplifying conditions of bar 2 being more massive than bar 1 and hence considered as rigid during
the interaction. In this situation, the system moves as if the bars were repelling each other, as 
far as their velocities are concerned, while adjusting their geometries. This happens because for
 each bar the parameter ratio governing the run of the rotational velocity turns out to be
$(2\Omega /a_1)( \partial a_1/\partial \Omega) <-1$, and $\Omega (\partial q_2/\partial \Omega)<0$,
which are conditions sufficient to give rise to a variation of the angular velocity with opposite sign to
that of the torque, according to Equation (\ref{eq_ppal_defj}). The system develops a periodic oscillation, 
in the cases that we have explored, which is not damped, as far as we could see. It has to be
stated that the double bar system is in equilibrium during the evolution, as dictated by 
Eq. (\ref{equilibrio}), which is used to derive the evolution of the system. In other words, 
each bar follows its track under the condition of hydrostatic equilibrium, balancing the 
centrifugal force due to its rotation with the centripetal attraction of the gravity and 
adjusting its geometry to maintain its ellipsoidal figure throughout all the time.

This result is along the same line as the conclusions in \citet{Abramyan1986}, who consider 
the case of the interaction of a self-gravitating triaxial bulge, treated as Riemann S-ellipsoid,
with a system formed by a combination of bar, halo and disc. In that study, the gravitational 
attraction of the bulge on the bar, mainly because of the small size and mass attributed to bulge, 
is neglected and they treat the disc and halo as spheroids. They consider the motion of stars 
in the bulge following elliptic orbits with some oscillation frequency. When this frequency is 
null, they find that the figures of equilibrium are Jacobi ellipsoids. Interestingly enough, 
they show that the orbital motions are in a direct sense (following the rotation) in the bar, but
 retrograde in the bulge. In other words, according to \citet{Abramyan1986}, the angular 
momentum of the stars in the bulge will oppose the angular velocity of the rotation system, 
and the authors related this with the fact that bulges and bars are usually observed perpendicularly
 to each other. Our result indicates something similar but somewhat different. As we have not 
included the orbits of the stars, our systems are deformable solids without internal structure, 
and we have obtained an exchange of angular momenta between the two bars (bulge + bar in
\citet{Abramyan1986}) that results in opposing motions. In view of the conclusions in 
\citet{Abramyan1986}, we may tentatively consider that the inclusion of orbital motions
 will reduce the expansion/contraction of the bars, given that part of the angular momentum 
exchange will be invested in the increase/decrease of the angular velocity of the stars in
 the bar and bulge.

Our results are also in line, broadly speaking, with those of DS07. While the positions of 
the peaks and valleys of the oscillations with respect to the relative angle between bars 
are opposite, as could have been predicted because of the net effect of repulsion between bars 
in our model, the overall shape of the oscillatory motion of the lightest bar is quite similar 
in both models. Interestingly enough, DS07 also found discrepancies between their result and 
that of \cite{Heller2001a} on the same point, this latter therefore being coincident with ours.

We plan to extend the numerical simulations and models for the non-rigid case, 
including the case in which bar 2 varies its parameters along with those of bar 1. The 
non-homogeneity of the bars, while obeying certain prescriptions, could also be 
introduced. For subsequent work, it is also intended to take into account additional galactic
 components, which will introduce major changes in the evolution of the system including 
dynamical friction which would tentatively damp the oscillations shown in the present paper. 
Finally, as commented in section \ref{.sdef}, the balance between the relaxation time of the bars
 and the change rate of bar parameters due to the interaction must be addressed once a more 
complete picture of the model galaxy be built.

\section*{Acknowledgments}

This work has been partially funded by the Spanish MICINN under the Consolider-Ingenio 
2010 Program grant CSD2006-00070: First Science with the GTC (http://www.iac.es/consolider-ingenio-gtc). 
FGL and MLC are supported by the grant AYA2012-33211 of the Spanish Science Ministry.


\bibliographystyle{apj}
\bibliography{paper_2bars} 

\appendix
\section{Calculation of the torque between bars}
\label{app_torq}
For convenience, let us set the reference standard as a triaxial system that rotates 
with bar 1, the axes being aligned with the principal axes of the ellipsoid representing 
the bar. For the ease of calculation, an intermediate reference system that rotates with
 bar 2 is also considered, related to the bar in the same way as in the previous case. 
The $X$-axes of these systems form an angle $\Delta\Phi $, which varies with time, 
hence providing the rotation matrix between the two systems.

Solving Eq.\ (\ref{eq_torq}) numerically implies, in principle, a twofold triple integration, 
as it contains two integrals over the volumes of the two bars. An alternative and simpler 
way to calculate the same torque is by means of the gravitational potential ${\cal V}_2$ 
created by the bar 2 over the bar 1 given by Eq.\ (\ref{potencial}) and express the volume
 of integration in ellipsoidal coordinates,

\begin{equation}
\tau_1 (\Delta \Phi )=-\rho _1\int _{V_1}d\mathbf{x}_1\frac{\partial {\cal V}_2(\mathbf{x}_1)}
{\partial (\Delta \Phi )}
 \label{eqap2_torq}
\end{equation}
\[\begin{array}{l}
=\frac{9}{16\pi }\frac{GM_1M_2}{a_{11}^3a_{12}^3}\int _0^{a_{11}}dr_1 r_1^4\int _{-\pi /2}^{\pi /2}d\theta _1\cos ^3\theta _1 \\
\int _0^{2\pi }d\phi _1 \left[\frac{\partial A_2}{\partial (\Delta \Phi )}f_1^2+ \frac{\partial B_2}{\partial (\Delta \Phi )}f_2^2
+\frac{\partial C_2}{\partial (\Delta \Phi )}f_3^2+\right. \\
\left.+2(A_2-B_2)f_1f_2 \right]\\
f_1=\cos \Phi _1\cos \Delta \Phi+q_{21}\sin \Phi _1 \sin \Delta \Phi \\
f_2=-\cos \Phi _1\sin \Delta \Phi +q_{21}\sin \Phi _1\cos \Delta \Phi \\
f_3=q_{31}\tan \theta _1, \\
\end{array}\]
with $A_2$, $B_2$, $C_2$ given in Equations (\ref{potencial}).

\section{Periodicity of the rotation in the rigid bodies case} 
\label{app_per}
Let us consider two coplanar rotating bars in a plane with moments of inertia $I_1$ and 
$I_2$, and angular velocities $\Omega_1$ and $\Omega_2$. The total kinetic energy and angular 
momentum are then

\begin{equation}
\displaystyle
 K = \frac{1}{2}I_1 \Omega _1^2 + \frac{1}{2}I_2 \Omega _2^2
 \label{eqap_ecin1}
\end{equation}

\begin{equation}
 L = I\Omega = I_1 \Omega _1 + I_2\Omega _2 \quad {\mathrm{with}}\quad I = I_1 + I_2
 \label{eqap_mang}
\end{equation}

Defining $\Delta \Omega _i \equiv \Omega _i - \Omega$, we get

\begin{equation}
\displaystyle
 \Delta \Omega _1 = - \frac{{I_2 }}{{I_1 }}\Delta \Omega _2 \hfill \\
 \label{eqap_domg}
\end{equation}

Combining eqs\ \ref{eqap_ecin1} and \ref{eqap_domg}, the kinetic energy can be expressed as,

\begin{equation}
\displaystyle K = \frac{1}{2}I\Omega^2 + \frac{1}{2}\Delta \Omega _1^2 \left( I_1 + \frac{I_1^2 }{I_2 } \right)
 \label{eqap_ecin2}
\end{equation}
or, in terms of the relative frequency, defined as

\begin{equation}
\displaystyle
 \dot{\Delta\Phi} \equiv \Omega _2 - \Omega _1 = \left( 1 +\displaystyle\frac{I_2}{I_1} \right)\Delta \Omega _2
 \label{eqap_rfreq}
\end{equation}

\begin{equation}
 K = \frac{1}{2}I\Omega^2 + \frac{1}{2}J\dot{\Delta\Phi}^2 \quad {\mathrm{with}}\quad J = \frac{I_1 I_2^2 + I_1^2 I_2 }
{(I_1 + I_2 )^2 } \hfill \\
 \label{eqap_ecin3}
\end{equation}

Hence, from the conservation of the total angular momentum, with constant moments of 
inertia ($I, J$ and $\Omega$, constants), we conclude that $K= K(\dot{\Delta\Phi})$. 
Combining this with the conservation of energy,

\begin{equation}
 E = U(\Delta \Phi ) + K(\dot{\Delta\Phi}) = \mathrm{constant}
 \label{eqap_etot}
\end{equation}
it can be seen that at each relative position of the bar, given by $\Delta\Phi$, there is a 
corresponding $\dot{\Delta\Phi}$, in reality, two symmetric values around the same position 
of the reference bar, corresponding to the double sign of the square root in Eq. (\ref{eqap_ecin3}), 
and a single value of $\ddot{\Delta\Phi}$, given by Equation (\ref{eq_ppal_rig}). So, at fixed time periods, 
the system repeats the position and velocity, thus producing a periodic motion.

\end{document}